%% file: main.tex
\documentclass[twocolumn]{aastex63}

\let\tablenum\relax
\usepackage{siunitx, xspace, amsmath}
\graphicspath{{./}{figures/}}

\DeclareSIUnit\squaredegree{deg^2}
\DeclareSIUnit\as{as} % alternate arcsecond
\DeclareSIUnit\parsec{pc}
\DeclareSIUnit\erg{erg}
\DeclareSIUnit\year{yr}
\newcommand{\Msun}{\,\mathrm{M}_\odot}

\newcommand{\snecsm}{SNe Ia-CSM\xspace}
\newcommand{\sneia}{SNe Ia\xspace}
\newcommand{\galex}{\textit{GALEX}\xspace}
\newcommand{\hst}{\textit{HST}\xspace}
\newcommand{\swift}{\textit{Swift}\xspace}
\newcommand{\Halpha}{H$\alpha$\xspace}
\newcommand{\gphoton}{\textsc{gPhoton}\xspace}
\newcommand{\fcsm}{$f_\text{CSM}$\xspace}

\newcommand{\samplesize}{1080\xspace}
\newcommand{\countnondet}{1076\xspace} % sample size - 4
\newcommand{\injrecsize}{1003\xspace}

\received{April 5, 2021}
\revised{November 3, 2021}
\accepted{November 18, 2021}
\published{}
\submitjournal{ApJ}

\shorttitle{The Occurrence Rate of SNe Ia-CSM}
\shortauthors{Dubay et al.}

\begin{document}

\title{Late-Onset Circumstellar Medium Interactions are Rare:\\ An Unbiased \textit{GALEX} View of Type Ia Supernovae}

\author[0000-0003-3781-0747]{Liam O. Dubay}
\newcommand{\ifa}{Institute for Astronomy, University of Hawai`i, 2680 Woodlawn Drive, Honolulu, HI 96822, USA}
\altaffiliation{IfA 2020 REU Program Participant}
\affiliation{Department of Astronomy, The Ohio State University, 140 West 18th Avenue, Columbus, OH 43210, USA}
\affiliation{Whitman College, 280 Boyer Ave, Walla Walla, WA 99362, USA}
\affiliation{\ifa}

\author[0000-0002-2471-8442]{Michael A. Tucker}
\altaffiliation{DOE CSGF Fellow}
\affiliation{\ifa}

\author[0000-0003-3429-7845]{Aaron Do}
\affiliation\ifa
\author[0000-0003-4631-1149]{Benjamin J. Shappee}
\affiliation\ifa
\author[0000-0002-5259-2314]{Gagandeep S. Anand}
\affiliation{Space Telescope Science Institute, 3700 San Martin Drive, Baltimore, MD 21218, USA}
\affiliation\ifa

\correspondingauthor{Liam Dubay}
\email{liam.dubay@gmail.com}

%****************************************************************************

\begin{abstract}

Using ultraviolet (UV) light curves we constrain the circumstellar environments of 1080 Type Ia supernovae (SNe Ia) within $z<0.5$ from archival \textit{Galaxy Evolution Explorer} (\textit{GALEX}) observations. All SNe Ia are required to have pre- and post-explosion \textit{GALEX} observations to ensure adequate subtraction of the host-galaxy flux. Using the late-time \textit{GALEX} observations we look for the UV excess expected from any interaction between the SN ejecta and circumstellar material (CSM). Four SNe Ia are detected near maximum light and we compare the \textit{GALEX} photometry to archival data, but we find none of our targets show convincing evidence of CSM interaction. A recent \textit{Hubble Space Telescope} (\textit{HST}) survey estimates that $\sim6\%$ of SNe Ia may interact with distant CSM, but statistical inferences are complicated by the small sample size and selection effects. By injecting model light curves into our data and then recovering them, we constrain a broad range of CSM interactions based on the CSM interaction start time and the maximum luminosity. Combining our \textit{GALEX} non-detections with the \textit{HST} results, we constrain occurrence of late-onset CSM interaction among SNe Ia with moderate CSM interaction, similar to that observed in PTF11kx, to $f_\text{CSM}\lesssim5.1\%$ between $0-500$ days after discovery and $\lesssim2.7\%$ between $500-1000$ days after discovery at $90\%$ confidence. For weaker CSM interactions similar to SN 2015cp, we obtain limits of $\lesssim16\%$ and $\lesssim4.8\%$, respectively, for the same time ranges.

\end{abstract}


%****************************************************************************

\section{Introduction}
\label{sec:Introduction}

Type Ia supernovae (\sneia) are thermonuclear explosions of carbon-oxygen white dwarf stars \citep[C/O WDs;][]{Hoyle1960-nucleosynthesis}, and are typically classified based on the lack of H and He emission and the presence of strong Si II absorption in their spectra \citep{Filippenko1997-SNeSpectra}. \sneia are important for many fields of astrophysics: they are useful as standardizable candles \citep{Phillips1993}, and they played a leading role in the discovery of dark energy \citep{Riess1998-DarkEnergy,Perlmutter1999-DarkEnergy}. \sneia also influence the chemical evolution and distribution of metals in the universe \citep[e.g.,][]{Greggio1983-SNI-Rates,Wiersma2011-CosmicMetals}. However, the nature of their progenitor systems is not fully understood. In particular, there is an ongoing debate about the relative contributions from the single degenerate (SD) and double degenerate (DD) channels \citep[see][for reviews]{Maoz2014-SNeIa-Review,Livio2018-ProgenitorReview,Ruiter2020-ProgenitorReview}. The existence of multiple channels for producing \sneia could lead to systematic errors in SN Ia-calibrated distances if the relative contributions evolve with cosmic time \citep[e.g.,][]{Howell2011-IaCosmologyReview,dAndrea2011-HubbleResiduals}.

In the DD scenario, two WDs merge after an inspiral from a tight binary \citep{Iben1984-IaBinary,Webbink1984-DDModel,Pakmor2012-WDMerger} or a head-on collision \citep{Benz1989-CollisionalDD,Thompson2011-CollisionalDD}. The theoretical rate of WD mergers is consistent with the observed rate of \sneia \citep[e.g.,][]{Yungelson1994-WDMergerRate,Ruiter2009-SNeIaRates} and the lack of H and He emission in the spectra of normal \sneia is easily explained by the DD model. However, DD progenitor systems are difficult to detect even within the Milky Way \citep{Rebassa-Mansergas2019-BinaryWDs}, and the merger of two WDs may result in a high-mass WD or neutron star rather than a thermonuclear explosion \citep{Nomoto1985-WDMerger,Saio1998-WDMerger,Shen2012-WDMerger}. Despite these issues, in recent years the DD scenario has become the leading model for most SN Ia progenitors.

Conversely, a SD system consists of a WD and a close non-degenerate companion, such as a red giant, helium star, or main sequence star \citep{Whelan1973-SDModel,Nomoto1982-SDModel,Yoon2003-SDModel}. In most models the WD accretes matter from its companion and explodes once it nears the Chandrasekhar mass \citep{Whelan1973-SDModel}. Because the explosion only occurs when the WD reaches its maximum mass, the SD model can readily account for the homogeneity of normal \sneia. However, the presence of a close non-degenerate companion should produce observable signatures such as photometric irregularities in the early light curve as the ejecta impact the companion star \citep[e.g.,][]{Kasen2010-EjectaCollision,Boehner2017-CompanionImprints}, emission lines in nebular-phase spectra produced by material stripped from the donor star \citep[e.g.,][]{Wheeler2075-BinarySystems,Marietta2000-CompanionImpact,Pan2012-SDEjecta}, and radio emission from interaction with material carried by the stellar wind \citep[e.g.,][]{Chevalier1982-RadioEmissionSNeII,Chevalier1982-EjectaMediumInteraction}. Recent searches for these observational signatures have not found any conclusive evidence of a SD progenitor system \citep[e.g.,][]{Panagia2006-RadioEmission,Chomiuk2016-RadioEmission,Fausnaugh2019-EarlyIaLightCurves,Tucker2020-SNeIaSpectra}. Tycho G has been proposed as the surviving companion of SN 1572, also known as Tycho's SN \citep[e.g.,][]{Ruiz-Lapuente2004-TychoSN} and is supported by a recent kinematic study \citep{Ruiz-Lapuente2019-TychoSN}, but \citet{Shappee2013-DegenerateSurvivors} argue the star is not luminous enough. Other searches for surviving non-degenerate companions have come up short \citep[e.g.,][]{Schaefer2012-ExCompanionSNR,Do2021-Progenitor1972E}.

While some or even most \sneia may result from the DD channel, some peculiar \sneia are more consistent with a SD progenitor system. One such subset of \sneia are those with evidence for a dense circumstellar medium (CSM) in close proximity to the explosion. These events, termed ``\snecsm'', are often more luminous and feature strong H emission lines \citep{Silverman2013-SNeIa-CSM}. The first two members of this class were SN 2002ic \citep{Hamuy2003-SN2002ic,Deng2004-SN2002ic,Kotak2004-SN2002ic,Wang2004-SN2002ic,Wood-Vasey2002-SN2002ic} and SN 2005gj \citep{Aldering2006-SN2005gj,Prieto2007-SN2005gj}. While some have argued that a core-collapse progenitor better explains these events \citep{Benetti2006-SN2002ic,Trundle2008-SN2005gj}, \citet{Fox2015-CSMProgenitors} found that late-time spectra of \snecsm are more consistent with a thermonuclear explosion. PTF11kx was the first unambiguous case of an SN Ia interacting with a dense CSM \citep{Dilday2012-PTF11kx,Silverman2013-PTF11kx}, and since its discovery, the list of unambiguous \snecsm has steadily grown \citep[e.g.,][]{Silverman2013-SNeIa-CSM,Yao2019-ZTF2018,Graham2019-SN2015cp,Srivastav2021-ePESSTO-spec-class}.

The presence and strength of CSM interaction can constrain the SN Ia progenitor system. A DD collision involving two C/O WDs may produce CSM, but with such a small H mass fraction ($M_{\rm{H}} / M_{\rm{WD}} \lesssim 10^{-4}$; \citealt{Romero2012-WDMass}) the amount of hydrogen ejected would be negligible. A He + C/O WD system is expected to eject $3-\SI{6e-5}{\Msun}$ prior to the merger \citep{Shen2013-DD-CSM},
which is likely too little mass to explain the \Halpha emission observed in \snecsm. By contrast, the mass-transfer process in the SD scenario is inefficient and may produce up to several $\Msun$ of H-rich CSM as material expelled by the companion (e.g., by wind from a red giant) is swept up by a nova eruption to produce a dense circumstellar shell \citep[e.g.,][]{Hamuy2003-SN2002ic,Walder2008-RSOphiuchi,Moore2012-RecurrentNovae}. Symbiotic progenitor systems in particular are expected to have a mass-loss rate of $\dot M\gtrsim 1.7\times 10^{-8} \Msun\,\mathrm{yr}^{-1}$, assuming wind velocity $v_w\sim \SI{100}{\kilo\meter\per\second}$ \citep{Hachisu1999-SymbioticProgenitors,Lundqvist2020-RadioConstraints}. \citet{Aldering2006-SN2005gj} estimate a mass of $\gtrsim 10^{-2} \Msun$ would be necessary to explain the observed \Halpha luminosity of SN 2005gj. While the SD scenario has trouble accounting for the lack of nebular \Halpha emission in most \sneia \citep[e.g.,][]{Leonard2007-IaNebularSpectra,Shappee2013-Nebular2011fe,Shappee2018-Nebular2012cg, Tucker2020-SNeIaSpectra}, it is a promising progenitor channel for \snecsm \citep{Silverman2013-SNeIa-CSM}.

Most \snecsm are discovered before or near maximum light and show evidence of CSM interaction within days of peak SN brightness. For example, a strong \Halpha line was present in the spectrum of SN 2002ic at +6 days past maximum light \citep{Hamuy2003-SN2002ic}, and \Halpha was visible in SN 2005gj before peak brightness \citep{Aldering2006-SN2005gj}. However, a handful of \snecsm have recently been discovered where there is a clear lack of CSM interaction in early observations, with the \Halpha emission appearing weeks or months later. The first example of a late-onset SN Ia-CSM was PTF11kx, which featured prominent H and Ca emission starting 59 days after explosion \citep{Dilday2012-PTF11kx} and persisting after +3.5 yr \citep{Silverman2013-PTF11kx,Graham2017-PTF11kx}. Additionally, time-variable Na absorption has been linked to the presence of CSM \citep{Sternberf2011-NaAbsorption} and may indicate an unusual geometry of the CSM \citep{Simon2009-VariableNaAbsorption}, which could also be associated with time-variable \Halpha emission. \citet{Dilday2012-PTF11kx} propose that PTF11kx resulted from a symbiotic nova progenitor, though \citet{Soker2013-CoreDegenerate11kx} offer an alternative explanation in the violent prompt merger scenario.

There is reason to expect that CSM interaction may begin even later in other \sneia. CSM shells generated by recurrent novae may reach $\sim\SI{e17}{\centi\meter}$ by the time of the next eruption \citep{Moore2012-RecurrentNovae}. At this distance, ejecta traveling at $\sim\SI{23000}{\kilo\meter\per\second}$ \citep[e.g.,][]{Garavini2005-SN1999ac} would not begin to interact for $\sim500$ days. Even current radio observations do not provide meaningful constraints on distant CSM shells \citep{Harris2021-RadioConstraints}. If CSM is often present at such a large distance from the WD, then typical SN observations might systematically miss its signatures, as they usually continue for only a few months after the explosion \citep[e.g.,][]{Hicken2012-CfA4}.

Intrinsic differences may also exist among the SNe Ia-CSM class itself. Most \snecsm occur in star-forming host galaxies, exhibit \Halpha luminosities of $L_{\rm{H\alpha}}\approx \SI{e40}{\erg\per\second}$, and have bright, slowly-evolving light curves \citep{Silverman2013-SNeIa-CSM}. ASASSN-18tb/SN 2018fhw was the first sub-luminous, fast-declining SN Ia observed to have \Halpha emission after maximum light \citep{Kollmeier2019-SN2018fhw,Vallely2019-ASASSN18tb}. ATLAS18qtd/SN 2018cqj, another low-luminosity and fast-declining event, showed \Halpha emission in spectra taken at $+193$ and $+307$ days after peak \citep{Prieto2020-SN2018cqj}. The \Halpha luminosity observed in ASASSN-18tb and ATLAS18qtd was much lower than in other known \snecsm and was inconsistent with both material stripped from a companion in a SD system \citep{Marietta2000-CompanionImpact,Liu2012-EjectaInteractionSimulations,Boehner2017-CompanionImprints} and typical \snecsm \citep{Tucker2020-ATLAS18qtd}. This complicates our understanding of \snecsm, as it is unclear whether these objects represent the extreme end of a continuous distribution or constitute a new class of thermonuclear explosions.

While \sneia are predominantly optical phenomena \citep{Filippenko1997-SNeSpectra,Brown2010-IaUV}, CSM interaction produces ultraviolet (UV) emission which distinguishes these events from both the underlying emission from the ejecta and their host-galaxy \citep[e.g., SN 2005gj;][]{Immler2005-SN2005gj}. To search for late-onset CSM interaction, defined as $\ge100$ days after peak brightness, \citet{Graham2019-SN2015cp} performed a \textit{Hubble Space Telescope} (\hst) near-ultraviolet (NUV) snapshot survey targeting 72 nearby \sneia $1-3$ yr after explosion. ASASSN-15og showed early signs of CSM interaction \citep{Monroe2015-ASASSN-15og,Holoien2017-ASASSN-DR2} and was detected in the NUV at +477 days after maximum light \citep{Graham2019-SN2015cp}. \citet{Graham2019-SN2015cp} also detected NUV emission in SN 2015cp at +664 days. SN 2015cp was originally classified as a SN Ia-91T, and showed no signs of CSM interaction in its spectrum at +45 days \citep{Frohmaier2016-PESSTOclass}, but subsequent spectra taken between +694 and +785 days revealed declining \Halpha and Ca II emission consistent with interaction between the SN ejecta and a distant shell of H-rich CSM \citep{Graham2019-SN2015cp}. This discovery demonstrates that late-onset \snecsm may be missed in typical SN observations. 

\snecsm are rare \citep{Silverman2013-SNeIa-CSM,Graham2019-SN2015cp}, but the true occurrence rate is not well constrained. \citet{Graham2019-SN2015cp} estimated that the fraction of their targets that have CSM within $r_\text{CSM}\approx\SI{3e17}{\centi\meter}$ is $f_\text{CSM}\approx6\%$. However, they selected targets with characteristics typical of \snecsm, such as an SN 1991T-like spectrum \citep[e.g.,][]{Phillips1992-SN1991T}, high photospheric velocity, a blueshifted Na I D absorption line, or a host with a young stellar population. Therefore, their sample is already biased towards finding \snecsm.

To better constrain the fraction of \sneia with late-onset CSM interaction, we search for UV emission from known \sneia in archival data from the \textit{Galaxy Evolution Explorer} spacecraft \citep[\galex;][]{Martin2005-GALEX}. In Section \ref{sec:TargetSelAndObs}, we describe our target selection and \galex observations. In Section \ref{sec:PhotometricAnalysis}, we present detections of normal \sneia and convert non-detections to limits on intrinsic UV luminosity. In Section \ref{sec:OccurrenceRate}, we constrain the occurrence rate of late-onset \snecsm. We present our conclusions in Section \ref{sec:Conclusions}. Throughout this work, we adopt  $H_0=\SI{70}{\kilo\meter\per\second\per\mega\parsec}$, $\Omega_m=0.3$, and $\Omega_\Lambda=0.7$. We present all observation times in terms of days after discovery in the SN rest frame.

%****************************************************************************

\section{Observations and Target Selection}
\label{sec:TargetSelAndObs}

We obtained \galex \citep{Martin2005-GALEX} UV light curves of \samplesize \sneia to search for signatures of SN Ia ejecta interacting with nearby CSM. \galex was a NASA Small Explorer telescope which surveyed the entire sky in the UV from 2003 to 2013, and its data are publicly available at the Mikulski Archive for Space Telescopes (MAST)\footnote{\url{https://archive.stsci.edu/}}. \galex is particularly suited for this purpose due to the low background noise of its photon-counting detectors \citep{Martin2005-GALEX} and the low surface brightness of SN Ia host galaxies in the UV \citep[e.g.,][]{depaz2007}. Previous studies have searched for UV emission from Type II SNe in \galex data \citep[e.g.,][]{Gal-Yam2008-SN2005ay, Gezari2008-TypeIIP, Gezari2010-SN2010aq, Gezari2015-PS1-13arp,  Ganot2016-GALEXSNeII, Soumagnac2019-PTF12glz, ganot2020}, but it has not yet been used for a large study of \sneia. 

In Section \ref{ssec:SurveyConfig}, we briefly describe the \galex spacecraft. In Section \ref{ssec:PhotometricStability}, we describe the data pipeline, and we address the photometric precision and stability of \galex. In Section \ref{ssec:Targets}, we describe our selection of a Type Ia sample. In Section \ref{subsec:nearby-historical}, we discuss very nearby ($z<0.01$) \sneia without pre-explosion imaging. We discuss our host-galaxy subtraction process in Section \ref{subsec:hostgal-subtraction}.

\subsection{Survey Configuration}
\label{ssec:SurveyConfig}

\galex operated in low-Earth orbit with a 50 cm objective and a \ang{1.2} circular field of view \citep{Martin2005-GALEX}. It obtained simultaneous images in the FUV ($1340-\SI{1800}{\angstrom}$) and NUV ($1700-\SI{3000}{\angstrom}$) bands until the FUV detector failed in 2009, after which the NUV detector operated alone until the end of the mission in 2013. \galex performed several imaging surveys during its decade of operation, including the All-Sky Imaging Survey (AIS), the Medium Imaging Survey (MIS), the Deep Imaging Survey (DIS), and the Nearby Galaxies Survey \citep[NGS;][see their Table 2]{Martin2005-GALEX}. The spacecraft covered nearly 77\% of the sky over its ten-year lifetime in at least one band \citep{Million2016-gPhoton} to a sensitivity of $\ge20.5$ AB magnitudes \citep{Martin2005-GALEX}.

Figure \ref{fig:FilterCurves} compares the \galex NUV and FUV filters to the \hst F275W filter and the \swift/UVOT filters. Filter response curves were provided by the Spanish Virtual Observatory (SVO) Filter Profile Service \citep{Rodrigo2012-SVO,Rodrigo2020-SVO}. The \galex NUV filter has an effective wavelength $\lambda_\text{eff} = \SI{2305}{\angstrom}$ and an equivalent width $W_\text{eff} = \SI{770}{\angstrom}$ which is similar to, but slightly wider than, the \swift UVM2 filter. The NUV filter is bluer and wider than the \hst F275W filter utilized by \citet{Graham2019-SN2015cp} in their search for \snecsm. The \galex FUV filter, with $\lambda_\text{eff} = \SI{1550}{\angstrom}$ and $W_\text{eff} = \SI{265}{\angstrom}$, has no direct \swift counterpart, covering shorter wavelengths than any of the UVOT filters. 

\begin{figure}[tb]
    \centering
    \includegraphics[width=\linewidth]{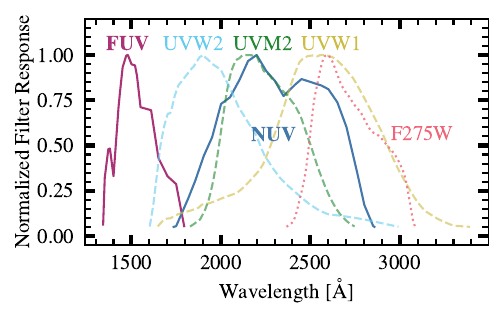}
    \caption{Filter response curves for \galex, \swift UV, and relevant \hst UV filters. The \galex NUV filter is comparable to the \swift UVM2 filter, and both \galex bands cover shorter wavelengths than the \hst F275W filter used by \citet{Graham2019-SN2015cp}.}
    \label{fig:FilterCurves}
\end{figure}

\subsection{Data Reduction and \gphoton Photometry}
\label{ssec:PhotometricStability}

We use the \gphoton package version 1.28.9 \citep{Million2016-gPhoton} to query \galex data products for our targets. No coadding is implemented to maximize temporal coverage, so single-epoch exposure times range from $\sim 100-1500~\rm s$. \citet{Million2016-gPhoton} found the relative astrometry between source positions in the \galex Merged Catalog (MCAT) and centers-of-brightness determined by \textsc{gAperture} to be better than $0\farcs01$, so we do not further correct the \galex astrometry.

Light curves are queried with an aperture radius of $6\arcsec$, equivalent to the \textsc{aper4} value in MCAT, and a background estimation annulus from $10\arcsec$ to $15\arcsec$. Choosing an aperture slightly larger than the image FWHM of $5\farcs5$ is a good compromise between capturing flux in the extended wings of the PSF and preventing background sources from contaminating the photometry \citep{Morrissey2007-GALEXcalib}. We use the background flux computed from the \gphoton background annulus instead of the MCAT-derived values, as the latter requires a nearby MCAT source for each observation to estimate the background flux. For some short \galex exposures or faint SN Ia host-galaxies no MCAT entries are available, resulting in undefined background flux levels. Using the aperture photometry method allows us to carry out a homogeneous analysis of our full sample.

\gphoton produces quality flags for the output light curve in several situations. Some flags are generated as ``warnings,'' whereas some flags are considered ``fatal'' and the photometry should not be trusted. Fatal flags include ``(bkgd) mask edge,'' ``exptime,'' ``nonlinearity,'' and ``spacecraft recovery.'' We exclude all light curve data with these fatal flags and refer the reader to the \gphoton documentation\footnote{\url{https://gphoton.readthedocs.io/en/latest/}} for their descriptions.

The ``detector response'' flag is the most common flag, affecting $\sim 25\%$ of the \galex data. This flag is set if any photon event falls outside $> 0.5~\rm{deg}$ from the center of the detector \textit{at any point} in the exposure. We find the photometry does not show a significant deviation from the MCAT magnitudes until $\gtrsim 0.6~\rm{deg}$ from the detector center. To improve our photometric completeness, we include photometry within $0.6~\rm{deg}$ from the detector center, increasing the effective area of the \galex detector by $\sim45\%$ compared to the nominal cut at $0.5~\rm{deg}$. A full description of our photometric testing and validation is provided in Appendix \ref{app:Photometry}.

All photometry is corrected for foreground Milky Way extinction using the dust maps of \citet{schlafly2011} and a \citet{cardelli1989} reddening law. This results in total-to-selective extinction values of $R_{\rm{NUV}}=7.95$ and $R_{\rm{FUV}} = 8.06$ (see Table 2 from \citealp{Bianchi2011-StarFormation}), although we caution that \citet{yuan2013} find a lower $R_{\rm{FUV}} \approx 4.5$ but a similar $R_{\rm{NUV}}$. We assume the $R_{\rm{FUV}}$ value from \citet{Bianchi2011-StarFormation} but note that the vast majority of \sneia are away from the Galactic plane and have little Galactic reddening, reducing the consequences of any discrepancy on $R_{\rm{FUV}}$.

\subsection{Target Selection}
\label{ssec:Targets}

We query the Open Supernova Catalog\footnote{\url{https://sne.space/}} \citep[OSC;][]{Guillochon2017-OSC} for objects classified as ``SN Ia'' and discovered prior to 2014 (as \galex was decommissioned in June 2013), returning 7265 objects. Several cuts are applied to the sample to reduce the number of non-\sneia objects contaminating our sample, prioritizing purity over completeness. Any objects with disputed classifications in the OSC (i.e., ``SN Ia, SN Ib/c'') are removed. Spectroscopic classification is considered robust, so any objects with only a ``SN Ia'' designation (or variant therein, e.g., ``Ia-91T-like'') and at least one publicly available spectrum are included in our sample. For objects designated ``SN Ia'' without a publicly available spectrum, we cross-match the SN Ia names with archival International Astronomical Union Circulars (IAUCs), Central Bureau Electronic Telegrams (CBETs), and Astronomer's Telegrams (ATels) to search for classification reports without publicly-released spectra.

Finally, we include photometrically-classified SN Ia from major photometric surveys including the Sloan Digital Sky Survey \citep[SDSS; ][]{York2000-SDSSLegacy} supernova survey \citep{sako2008, sako2011, Sako2018-SDSS-II}, the Panoramic Survey Telescope and Rapid Response System \citep[Pan-STARRS; ][]{chambers2016, jones2017, jones2018}, the SuperNova Legacy Survey \citep[SNLS; ][]{guy2010, conley2011, sullivan2011}, and the ESSENCE supernova survey \citep{Miknaitis2007-ESSENCE, woodvasey2007, narayan2016}. Photometrically-classified \sneia are required to have $P_\text{Ia} \geq 0.99$, where $P_\text{Ia}$ is the probability the transient is a SN Ia (see, e.g., \citealp{sako2011}). If multiple surveys observed and classified the same object, we use the highest reported $P_\text{Ia}$ value. This probability is used only  to determine which \sneia to include in our sample and is not used to weight our results.

\begin{figure}[tb]
    \centering
    \includegraphics[width=\linewidth]{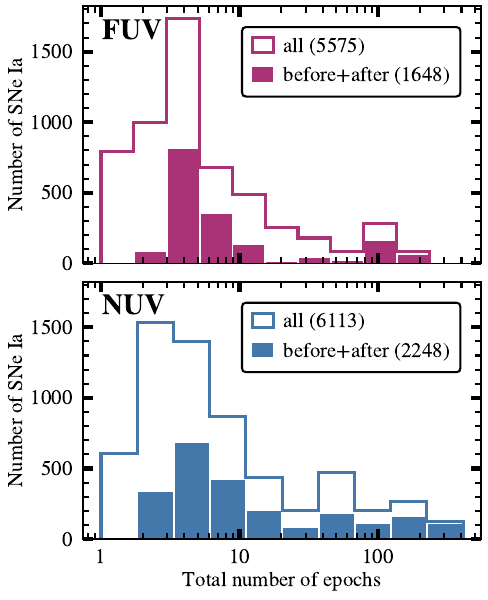}
    \caption{The number of \sneia that were observed for a given number of epochs by \galex. Outlined bars include all \sneia with at least a single epoch, while filled bars represent those with observations both before and after the date of discovery. Because the FUV detector failed in 2009, there are more NUV than FUV images and some \sneia have only NUV images.}
    \label{fig:ObservationHist}
\end{figure}

\begin{figure}[bt]
    \centering
    \includegraphics[width=\linewidth]{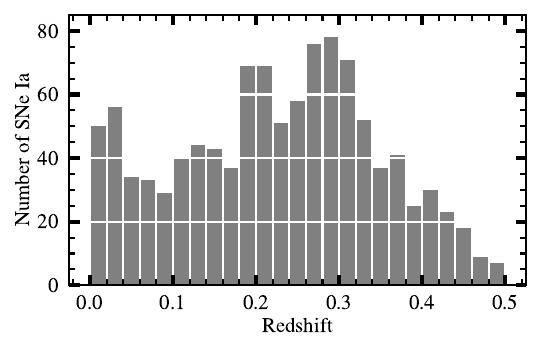}
    \caption{The redshift distribution of the \samplesize \sneia.}
    \label{fig:Redshifts}
\end{figure}

Additional cuts are applied to ensure the \galex observations are of sufficient coverage and depth. A cut of $z \leq 0.5$ is applied to ensure detectable emission, as a \galex single-visit limiting magnitude of $\sim 22.5$~mag corresponds to an absolute magnitude of $\sim -19.5$~mag at $z=0.5$. Additionally, all \sneia are required to have pre- and post-discovery \galex observations to ensure adequate host-galaxy subtraction. As Figure \ref{fig:ObservationHist} shows, 2248 \sneia were observed by \galex both before and after discovery ($t_\text{disc}$), of which all had NUV data and 1648 also had FUV coverage.

% Redshifts and cuts
\sneia with sufficient \galex coverage and $z\leq0.5$ are cross-matched with galaxies in the NASA/IPAC Extragalactic Database (NED)\footnote{\url{https://ned.ipac.caltech.edu/}} to check for more precise redshifts. \sneia with insufficient redshift precision are discussed in Appendix \ref{app:manualz}. We require a projected distance of $<\SI{100}{\kilo\parsec}$ between the SN Ia location and the center of the host-galaxy to prevent spurious matches and flag \sneia with projected offsets of $\geq\SI{30}{\kilo\parsec}$ for manual review. Figure \ref{fig:Redshifts} shows the final distribution of redshifts in our sample of \samplesize \sneia.

% Distances and estimated uncertainties
We also obtain high-precision values for redshift-derived distance and Milky Way extinction from NED. To account for the effect of galactic peculiar velocity, we add an additional systematic distance error of $\SI{300}{\kilo\meter\per\second}/H_0$ \citep{Zaroubi2002-CosmicFlows,Karachentsev2006-PeculiarVelocities} to the uncertainty in the redshift-derived distance estimates. Finally, we incorporate high-quality redshift-independent distances from the \textit{Cosmicflows-3} catalog \citep{Tully2016-Cosmicflows3} where available. Table \ref{tab:Targets} presents a subset of our sample.

% Nearby historical SNe
\subsection{Nearby, Historical SNe Ia}
\label{subsec:nearby-historical}

There are a number of \sneia which were discovered before \galex launched and have extensive post-discovery coverage. We identify 104 \sneia with $z < 0.01$ which were observed by \galex only after discovery. Of these, 13 \sneia had at least 10 epochs in at least one band. We provide a list of these targets in Table~\ref{tab:nearby-historical}.

Visual inspection of the \galex light curves reveals no obvious excess flux after the near-peak epoch ($t_\text{disc}<50$~days). At a distance of $\SI{10}{\mega\parsec}$, \galex should be sensitive down to an absolute magnitude of $M_{\text{UV}} \sim -7.5$ mag, and at $\SI{20}{\mega\parsec}$ it should be sensitive down to $M_{\text{UV}} \sim -9$ mag. Typical \snecsm have peak absolute magnitudes of $-21.3 \le M_{\text{R}} \le -19$ mag \citep{Silverman2013-SNeIa-CSM}, and \citet{Graham2019-SN2015cp} detected NUV emission from SN 2015cp at $M_{\text{F275W}}=-13.1$ mag hundreds of days after peak brightness. While these nearby, historical \sneia have high-quality limits from \galex, we exclude them from our statistical analysis because they lack pre-explosion imaging necessary for host-galaxy subtraction.

\subsection{Host-Galaxy Subtraction}
\label{subsec:hostgal-subtraction}

Our sources are restricted to \sneia with \galex observations both before and after discovery to ensure adequate subtraction of the host-galaxy flux. This eliminates $\sim3900$ \sneia with only pre- or post-explosion \galex observations. Normal \sneia have a $B$-band rise time between explosion and maximum brightness of $\sim16-25$ days \citep{Ganeshalingam2011-SNIaRiseTimes,Firth2015-RisingLightCurves}, while \snecsm have somewhat longer rise times in the range of $\sim20-40$ days \citep{Silverman2013-SNeIa-CSM}. The OSC only reports the discovery date, so we use $t_0=t_\text{disc}-30$~days as a conservative estimate for the date of explosion to avoid including any SN flux in our background measurements.

We use two methods to estimate the host galaxy flux depending on the number of pre-SN observations available. For \sneia with $\geq 5$ \galex observations prior to discovery, we compute the weighted average of all single-epoch pre-discovery fluxes and use the weighted standard deviation to estimate the associated uncertainty. We expand the formal statistical uncertainty by adding a systematic error contribution in quadrature until the fit has a $\chi^2$ per degree of freedom of unity, and then use these revised uncertainties to compute the uncertainty in the mean.

When there are $<5$ \galex observations prior to the SN discovery date in a given filter it is more difficult to empirically calibrate the uncertainties. If there are multiple pre-discovery observations, we compute the weighted mean and standard deviation of the single-epoch fluxes similar to the process described above. Then, we include an additional magnitude-dependent systematic uncertainty in quadrature which is described in Appendix \ref{app:host-systematics}. After computing the host-galaxy flux and associated uncertainty, the host-galaxy fluxes are subtracted from the post-discovery \galex observations.

%****************************************************************************

\section{Photometric Analysis}
\label{sec:PhotometricAnalysis}

We flag targets for review if the host-subtracted light curve has at least one detection at $\ge 5\sigma$ significance or at least three detections at $\ge 3\sigma$ significance. Out of

\input{short_sample.tex}

\input{nearby_historical.tex}

\noindent the \samplesize \sneia in our sample, 10 are flagged for review. One is detected in just the FUV band and the rest have only NUV detections. We reject three candidates with faint host-galaxies because the uncertainties in the host flux appear to be underestimated. All have few ($2-3$) host measurements with low signal-to-noise ratios. For $3\sigma$ significance and assuming Gaussian-distributed uncertainties, we expect $1-2$ host-galaxies to have underestimated host-galaxy fluxes which is consistent with the three false positives flagged by our search algorithm. A fourth candidate is rejected because the flagged image frame showed significant ghost artifacts. We eliminate a fifth candidate which appears to have an underestimated background flux because the 5 epochs within $\pm3$ days of the flagged frame are not significantly above the host-galaxy and no obvious source is visible in the flagged image frame. One other candidate, ESSENCEn263, has UV detections consistent with the center of the host galaxies and is likely an AGN flare, which we discuss in Appendix \ref{app:false-detect}.

\subsection{Detections of Normal \sneia}
\label{ssec:NormalDetections}

The four remaining candidates, all detected in the NUV, are detections of normal \sneia near maximum light. Figure \ref{fig:NormalSNe} shows the near-peak light curves of SN 2007on \citep[see][]{Pollas2007-SN2007on}, SN 2008hv \citep{Pignata2008-SN2008hv}, SN 2009gf \citep{Nakano2009-SN2009gf}, and SN 2010ai \citep{Caldwell2010-SN2010ai}. The \galex NUV light curves for SNe 2007on, 2008hv, and 2009gf are consistent with the \swift UV light curves from \citet{Brown2014-SOUSA}. There are no UV observations of SN 2010ai in the literature, so we present its \galex NUV light curve alongside optical measurements by \citet{Hicken2012-CfA4}. For all photometric data we assume a monochromatic flux density as in the AB magnitude system \citep{Oke1983-StandardStars}.

\begin{figure*}[tb]
    \centering
    \includegraphics[width=\textwidth]{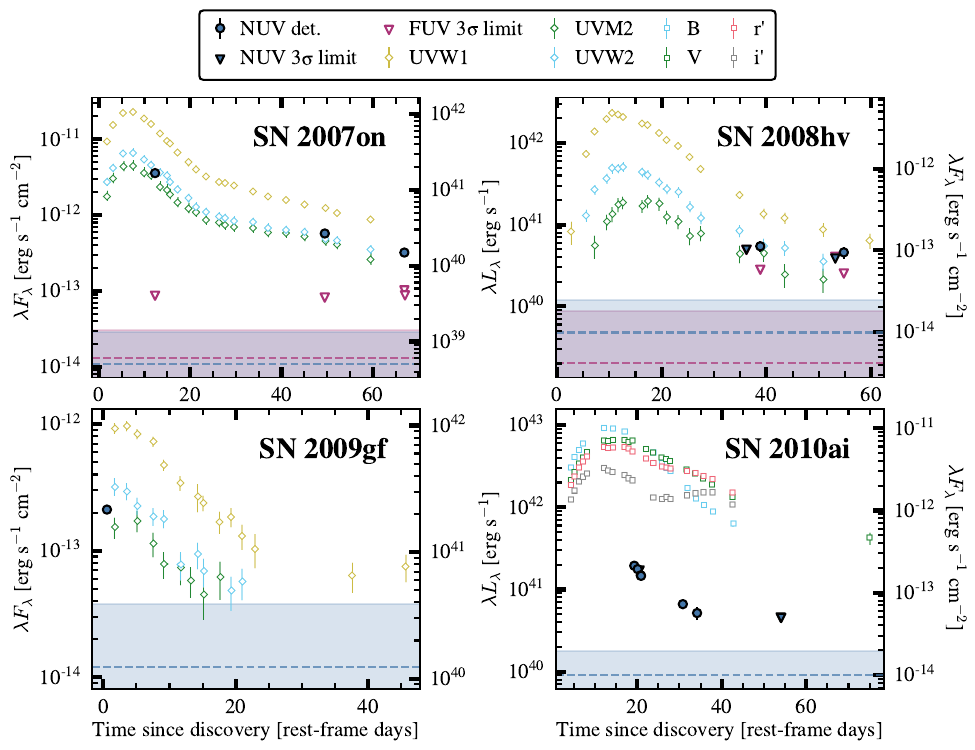}
    \caption{\galex light curves of the four \sneia detected near maximum light. The blue circles represent $\ge 3\sigma$ detections in the NUV, and the inverted pink and blue triangles represent $3\sigma$ non-detection limits in the FUV and NUV, respectively. Dashed lines and shaded regions represent the host-galaxy flux and associated $1\sigma$ uncertainty, respectively. The number of pre-discovery observations for each SN Ia in the NUV (FUV) are 34 (26) for 2007on, 3 (4) for 2008hv, 1 (0) for 2009gf, and 4 (0) for 2010ai. Near-peak \swift UV light curves from \citet{Brown2014-SOUSA} are included for SNe 2007on, 2008hv, and 2009gf. SN 2010ai does not have any other UV photometry available in the literature, so we include optical photometry from CfA4 \citep{Hicken2012-CfA4}. The interior vertical axis converts observed flux $\lambda F_\lambda$ to luminosity $\lambda L_\lambda$, corrected for Milky Way extinction.
    \label{fig:NormalSNe}}
\end{figure*}

We can also constrain the near-peak FUV flux for SN 2007on and SN 2008hv. For SN 2007on, we constrain its FUV emission at $t_\text{disc}+12$ rest-frame days to be $< \SI{2.2e36}{\erg\per\second\per\angstrom}$ at $3\sigma$ confidence, or $\lesssim 3$\% of the NUV emission at that epoch. For SN 2008hv, we constrain its FUV emission at $t_\text{disc}+39$ days to be $< \SI{1.68e37}{\erg\per\second\per\angstrom}$, or $\lesssim 78\%$ of the NUV emission at that epoch. \citet{Sauer2008-SyntheticSpectra} found that the flux at 1500 \si{\angstrom} should be an order of magnitude lower than at 2250 \si{\angstrom} in their model of the UV spectrum for SN 2001ep (see their Figure 4). Our non-detections in the FUV are qualitatively consistent with this model.

None of the four near-peak \sneia are candidates for CSM interaction. SN 2007on has been identified as a ``transitional'' SN Ia, in between the SN 1991bg and normal classes of \sneia \citep{Gall2018-SN2007on}, and it shows no signatures of CSM in its nebular spectrum \citep{Mazzali2018-SN2007on-Nebular,Tucker2020-SNeIaSpectra}. We also report seven NUV non-detections for SN 2007on between $t_\text{disc}+724$ days and $t_\text{disc}+753$ days, where we constrain the NUV luminosity to $< \SI{1.28e36}{\erg\per\second\per\angstrom}$. \citet{Challis2008-SN2008hv} reported that a spectrum of SN 2008hv taken before maximum was consistent with a normal SN Ia, though \citet{Marion2008-SN2008hv} suggested it to be a high-velocity-expansion SN Ia \citep[see][]{Wang2008-HighVelocitySNIa}. SN 2009gf was found to be a normal SN Ia several days before maximum \citep{Somero2009-SN2009gf}. \citet{Caldwell2010-SN2010ai} likewise identified SN 2010ai as a normal SN Ia a few days before peak brightness. Of the four \sneia, only SN 2007on has \galex observations after $t_\text{disc}+60$ days, and none of our detections are of an unusually high UV flux which would indicate a potential instance of CSM interaction.

%****************************************************************************

\subsection{Non-Detections \& Luminosity Limits}
\label{ssec:NonDetectionLimits}

We observe no evidence of CSM interaction in any of the \samplesize \sneia in our sample. All of the detections are either near-peak normal \sneia or detections of unrelated events. We can, however, convert our non-detections into limits on the intrinsic UV luminosity of the remaining \countnondet \sneia. We convert flux limits into intrinsic luminosity limits by using the distances listed in Table \ref{tab:Targets} and correcting for Milky Way extinction.

Figure \ref{fig:Limits} shows all post-discovery \galex data from our survey. Inverted triangles indicate $1\sigma$ upper limit non-detections, whereas filled points mark detections. For comparison, we also include \swift data of the normal SN Ia 2011fe \citep{Brown2012-11feSwift}; we use photometry from the UVM2 band because it aligns most closely with the \galex NUV filter profile (see Figure \ref{fig:FilterCurves}). We assume a monochromatic flux density for all observations. The FUV emission from \sneia is expected to be an order of magnitude lower than the NUV emission \citep{Sauer2008-SyntheticSpectra}, which is consistent with several FUV non-detections below the UVM2 light curve for SN 2011fe.

\begin{figure*}[tb]
    \centering
    \includegraphics[width=\textwidth]{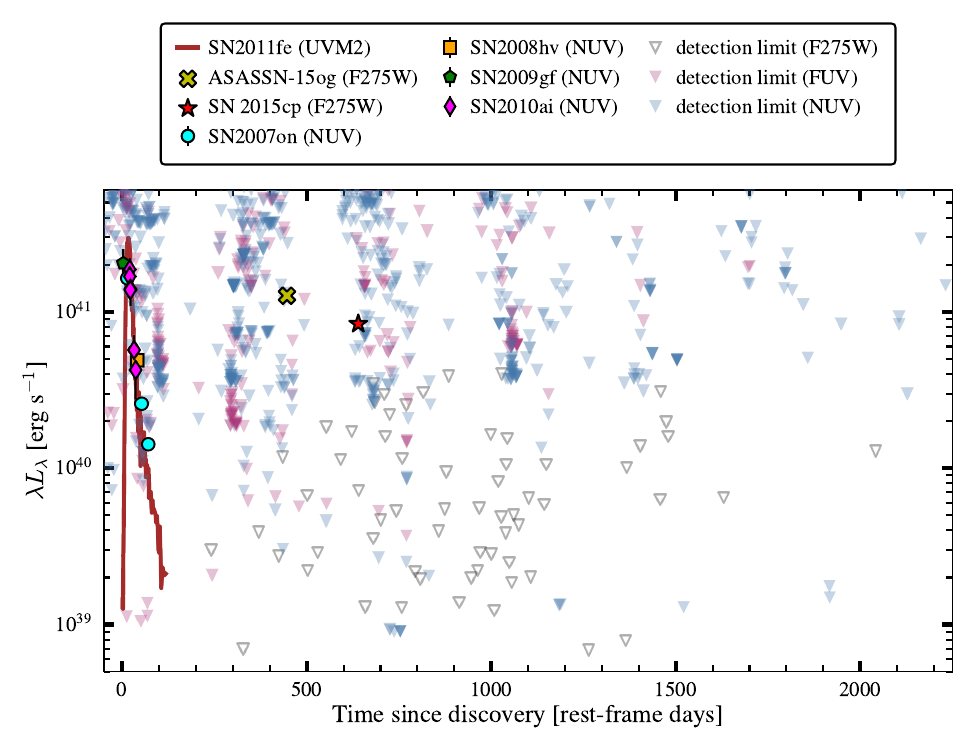}
    \caption{UV luminosities and limits from our \galex survey. Pink and blue inverted triangles represent $1\sigma$ non-detection limits in the FUV and NUV, respectively. Our near-peak \galex detections of \sneia 2007on (blue circle), 2008hv (orange square), 2009gf (green pentagon), and 2010ai (magenta diamond) are also shown. The solid brown line represents the \textit{Swift} UVM2 light curve of the normal SN Ia 2011fe \citep{Brown2012-11feSwift} for reference. The red star and yellow \textit{X} represent late-time \hst F275W detections of the \snecsm 2015cp and ASASSN-15og, respectively, by \citet{Graham2019-SN2015cp}, and the open gray inverted triangles represent their $1\sigma$ non-detection limits. All points are corrected for Milky Way extinction.}
    \label{fig:Limits}
\end{figure*}

Most of these non-detections do not rule out CSM interaction, especially at higher redshift. Many limits for \sneia at higher redshifts are too weak to eliminate even near-peak SN Ia flux. There are 66 \sneia with upper limits below $\lambda L_\lambda=\SI{8.4e40}{\erg\per\second}$, the luminosity of CSM interaction observed in SN 2015cp \citep{Graham2019-SN2015cp}. We include the 2 detections and 70 non-detections by \citet{Graham2019-SN2015cp} on Figure \ref{fig:Limits} for comparison.

The periodic nature of \galex observations present in Figure \ref{fig:Limits} is explained by the idiosyncrasies of SN surveys and the \galex spacecraft orbit. Similar to ground-based observations, \galex was restricted to pointing away from the Sun while observing. This created periodic gaps in the observing cadence, as a discovered SN is likely Sun-constrained 6 months after discovery and again 18 months after discovery, matching the data gaps seen in Figure \ref{fig:Limits}.

%****************************************************************************

\section{The Occurrence Rate of Late-Onset CSM Interaction}
\label{sec:OccurrenceRate}

We use our non-detection limits to constrain the fraction of \sneia which experience late-onset CSM interaction. To do this we assume a simple model for the \galex FUV and NUV and \hst F275W light curves of \snecsm which is described in Section \ref{ssec:EmissionModel}. In Section \ref{ssec:InjectionProcedure}, we describe the injection-recovery procedure to determine the number of \sneia in our sample which we exclude from showing signs of CSM interaction. We also run a similar procedure on the 72 \hst observations by \citet{Graham2019-SN2015cp}. In Section \ref{ssec:RecoveryResults}, we present the results of the recovery procedure on both data sets, which we use to constrain the fraction of \snecsm at multiple epochs in Section \ref{ssec:ConstrainingCSM}.

\subsection{CSM Emission Model}
\label{ssec:EmissionModel}

To interpret our UV non-detections, we require an understanding of how the emission properties of \snecsm evolve with time. Recent progress has been made in the radio regime \citep[e.g., ][]{Harris2016-CSMRadioModel, Harris2018-SN2015cp,Harris2021-RadioConstraints}, but there are presently no models for the UV emission or published UV spectra of \snecsm. The UV light curve model we adopt follows the same basic formulism as \citet{Graham2019-SN2015cp}. The ejecta encounters the CSM at time $t_{\rm{start}}$ days after explosion, producing an instantaneous rise in luminosity to $L_{\rm{max}}$. The luminosity remains constant at $L_{\rm{max}}$ for a plateau width of $W$ days, followed by a fractional decline in flux per 100 days $\Phi$. $L_{\rm{max}}$ and $t_{\rm{start}}$ are poorly constrained due to the small number of known late-onset \snecsm, but estimates for $W$ and $\Phi$ can be deduced from prior observations of \snecsm. We adopt a plateau width of $W=250$~days and $\Phi = 0.3$ to match the observations of PTF11kx \citep{Silverman2013-PTF11kx, Graham2017-PTF11kx}. While these parameters presumably vary over some range, these simple assumptions are necessary to reduce the total number of parameters.

As \galex observed in 2 filters (NUV and FUV) compared to the single \hst UV filter utilized by \citet{Graham2019-SN2015cp}, knowledge of the underlying spectral energy distribution (SED) is required to properly model the filter-specific luminosity. We use two simple models for the CSM emission: a flat-spectrum model and a line-emission model derived for Type II SNe interacting with nearby CSM.

The flat-spectrum model assumes a constant luminosity $L_\lambda$ for all filters (i.e., $L_{\rm{NUV}}(t) \equiv L_{\rm{FUV}}(t) \equiv L_{\rm{F275W}}(t)$). If the CSM emission is continuum-dominated, this approximation is adequate if the continuum is roughly blackbody and peaks in the UV, as was the case for SN 2005gj at early times \citep{Aldering2006-SN2005gj}. However, it is likely that \snecsm are only continuum-dominated in the earliest stages of CSM interaction.

The line-emission model is derived from the \citet{Chevalier1994-TypeIICSM} model spectrum for Type II SNe interacting with nearby CSM. Although it was developed for core collapse SNe, the physical processes governing the ejecta-CSM interaction are similar. In this model all the UV emission is due to lines as shown in Figure \ref{fig:Chev94Spectrum} for 1 year after explosion. The 1-year post-explosion model line ratios from \citet{Chevalier1994-TypeIICSM} agree well with the inferred emission-line ratios for SN 2015cp \citep{Graham2019-SN2015cp} and PTF11kx \citep{Dilday2012-PTF11kx, Silverman2013-PTF11kx}. The line width is assumed to be $2000~\rm{km}~\rm{s}^{-1}$, consistent with observations of \snecsm emission lines (mainly \Halpha) several months to years after maximum light \citep[e.g., ][]{Kotak2004-SN2002ic, Aldering2006-SN2005gj, Silverman2013-PTF11kx, Silverman2013-SNeIa-CSM, Graham2017-PTF11kx, Graham2019-SN2015cp}.

The line-emission model provides an avenue for probing \snecsm at higher redshifts than previous studies. To determine $L_\text{max}$, the spectrum is first redshifted to the SN Ia redshift and then integrated over the \galex or \hst filters. The Lyman-$\alpha$ emission line, the strongest emission line in the model by a factor of $\sim6$, enters the \galex FUV filter at $z\approx0.1$. Figure \ref{fig:Chev94FilterLum} shows the specific luminosity for each filter as a function of redshift, highlighting the importance of FUV observations for moderate-redshift \sneia.

\begin{figure}
    \centering
    \includegraphics[width=\linewidth]{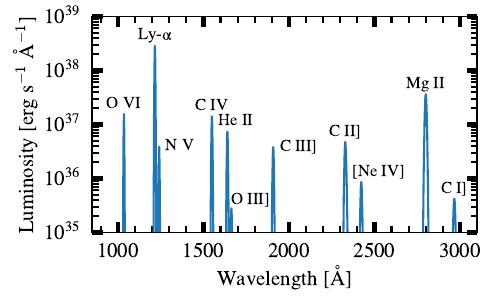}
    \caption{UV line emission model for CSM interaction in SNe II from \citet[see Table 6]{Chevalier1994-TypeIICSM}. Labels for the weaker lines Si II] (2335 \AA) and O V] (1218 \AA) are omitted for clarity.}
    \label{fig:Chev94Spectrum}
\end{figure}

\begin{figure}
    \centering
    \includegraphics[width=\linewidth]{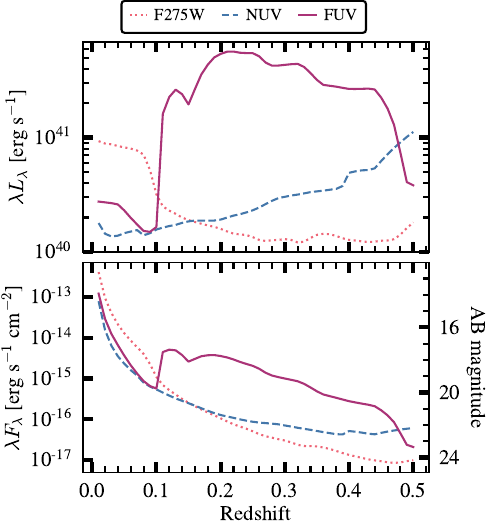}
    \caption{\textit{Top:} Luminosity of CSM emission using the line-emission spectral model for SNe II \citep{Chevalier1994-TypeIICSM} in \galex FUV and NUV bands and the \hst F275W band as a function of redshift. 
    \textit{Bottom:} Brightness of the line-emission model as a function of redshift and luminosity distance.}
    \label{fig:Chev94FilterLum}
\end{figure}

We use a dimensionless scale factor $S$ to calibrate the models to known \snecsm, where we define $S=1$ at the observed luminosity of CSM interaction in SN 2015cp, $L_\text{F275W}=\SI{3.1e37}{\erg\per\second\per\angstrom}$ at $z=0.0413$ \citep{Graham2019-SN2015cp}. On this scale PTF11kx is $S=19$ and SN 2005gj is $S=54$, as shown in Table \ref{tab:ReferenceScaleFactors}.

\input{scale_factors.tex}

\subsection{Injection Procedure}
\label{ssec:InjectionProcedure}

The model light curves are injected into the \galex and \hst data as early as 0 days after discovery. To eliminate contamination from near-peak UV emission, we only search for CSM emission for observations at $t>t_\text{disc}+50$ days, reducing the \galex sample to \injrecsize \sneia. The wide luminosity plateau allows models with early $t_\text{start}$ to be constrained by later observations. We also remove observations with $\ge3\sigma$ detections, whether they are near-peak detections, unrelated events, or spurious detections (see Section \ref{ssec:NormalDetections}). For objects in the \hst survey, we convert their 50\% limiting magnitudes \citep[see Table 3 in][]{Graham2019-SN2015cp} to $3\sigma$ upper limits on the intrinsic UV luminosity.

We apply an ``SED correction factor'' to the reported fluxes from \galex and \hst before injection. Because \gphoton reports monochromatic AB magnitudes \citep{Million2016-gPhoton}, it inherently assumes an SED which is flat in $F_{\nu}$. To make meaningful comparisons to the model CSM emission, especially the line-emission model, it is necessary to replace this assumed spectrum with the flat and line-emission models as a function of redshift. The correction factor is calculated by shifting the model spectrum by the redshift of the source, integrating over the given filter, and dividing by the flux of the AB magnitude zero point, $F_{\nu,\rm{zp}} = \SI{3.63e-20}{\erg\per\second\per\centi\meter\squared\per\hertz}$, integrated over the same filter. We set the scale so that the correction factor for an object with $z=0.0413$ (i.e., SN 2015cp) in the F275W band is equal to one.

We sample $t_\text{start}$ from a uniform distribution relative to $t_\text{disc}$ of $0-2500$ days, and we logarithmically sample $S$ over $0.01-100$. For each target, we randomly sample $N=10\,000$ instances of $t_\text{start}$ and $S$ from this parameter space. The generated model light curve is then injected into the target SN Ia light curve. If the injected signal reaches $\ge3\sigma$, that SN Ia is excluded from showing CSM interaction at those parameters. Targets in the \galex sample are excluded if the significance threshold is reached in either filter.

\subsection{Recovery Results}
\label{ssec:RecoveryResults}

Figure \ref{fig:ExcludedSNeIa} shows 2D histograms of the number of excluded \sneia as a function of $t_\text{start}$ and $S$. We also outline the model parameter space where the two \snecsm observed by \citet{Graham2019-SN2015cp}, ASASSN-15og and SN 2015cp, are recovered by the injection-recovery procedure. The actual \hst detections correspond to $S\approx1.5$ and $S=1$, respectively.

\begin{figure*}
    \centering
    \includegraphics[width=\textwidth]{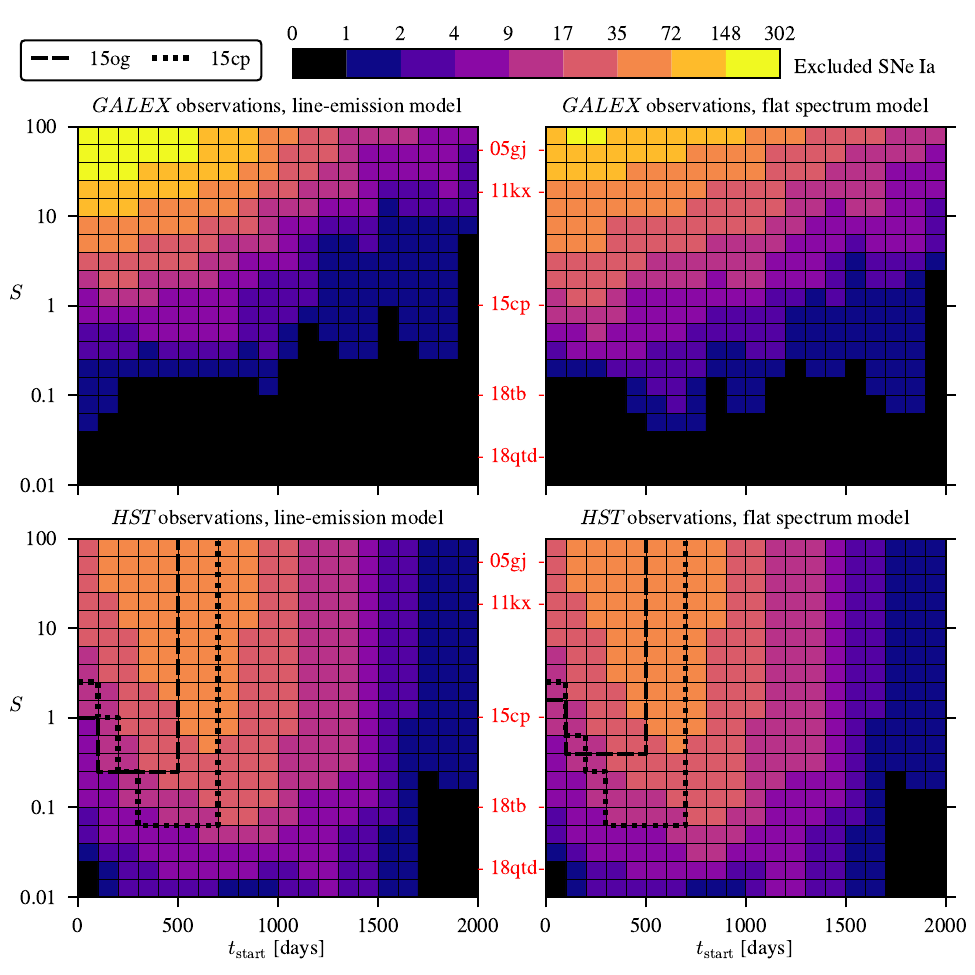}
    \caption{Recovery results showing the number of \sneia excluded from CSM interaction at varying start times and scale factors. \textit{Top:} \galex non-detections of \injrecsize \sneia in FUV and NUV bands. \textit{Bottom:} \hst observations by \citet{Graham2019-SN2015cp} of 72 \sneia in the F275W band. The parameter spaces where we recover ASASSN-15og and SN 2015cp (the \snecsm which were detected in the \hst survey) are outlined by dashed and dotted lines, respectively. Plots on the left use a line-emission model for SNe II \citep{Chevalier1994-TypeIICSM}, while plots on the right use a flat-spectrum model (see Section \ref{ssec:EmissionModel}). The scale factor is normalized so that unity corresponds to the UV luminosity of CSM interaction in SN 2015cp \citep{Graham2019-SN2015cp}. \galex data for the 50 days after discovery are omitted to avoid potential contamination from near-peak SN Ia flux, but model light curves are still injected starting from $t_\text{start}=0$ days post-discovery.}
    \label{fig:ExcludedSNeIa}
\end{figure*}

The horizontal axis of Figure \ref{fig:ExcludedSNeIa} bins $t_\text{start}$ in 100-day increments. We do not display results beyond $t_\text{disc}+2000$ days as very few \sneia could have been detected at such late times because of the $\sim3600$ day spacecraft lifetime combined with many SN surveys only starting after launch.

The scale factor $S$ on the vertical axis of Figure \ref{fig:ExcludedSNeIa} is binned into 20 logarithmic increments. Along the vertical axis we mark several known \snecsm as a proxy for the strength of the CSM emission. As only SN 2015cp was observed in the UV\footnote{SN 2005gj was observed in our sample at $>1000$ days after discovery, but it did not show significant UV flux.} \citep{Graham2019-SN2015cp}, we estimate $S$ using the observed \Halpha emission relative to SN 2015cp,

\begin{equation}
    \label{eq:HalphaRatio}
    S \equiv \frac{L_{\text{H}\alpha}}{L_{\text{H}\alpha}(\text{15cp})},
\end{equation}

\noindent and provide estimates for comparison \sneia in Table \ref{tab:ReferenceScaleFactors}. We also include similar ratios for two tentative \snecsm, ASASSN-18tb/SN 2018fhw \citep{Kollmeier2019-SN2018fhw,Vallely2019-ASASSN18tb} and ATLAS18qtd/SN 2018cqj \citep{Prieto2020-SN2018cqj,Tucker2020-ATLAS18qtd}, which showed weak \Halpha emission after peak brightness. Both events were sub-luminous \sneia so their scale factors are well below SN 2015cp.

The number of excluded \sneia in the \galex sample skews heavily to scale factors of $S\gtrsim10$, while the \hst sample has comparatively little dependence on $S$. This is partly a function of redshift because as the distance to the SN Ia increases, the minimum detectable CSM interaction luminosity also increases. The \galex sample, by definition, includes \sneia up to $z=0.5$ with an average of $\bar z\approx0.238$ (see Figure \ref{fig:Redshifts}), while the targets observed by \citet{Graham2019-SN2015cp} are much closer ($z\le0.08$) and more evenly distributed in redshift (see their Table 1). At large $S$ (i.e., SN 2005gj-like events), \galex can exclude many more \sneia than \hst due to the much larger sample size.

The \hst survey also has fainter flux limits than \galex. \citet{Graham2019-SN2015cp} report limiting AB magnitudes between 25.5 and 26 mag, while \galex has a limiting AB magnitude of $m_\text{F275W}\approx23.5$ mag for the Medium Imaging Survey or $\sim20.5$ mag for the All-Sky Survey \citep{Martin2005-GALEX}. This leads \citet{Graham2019-SN2015cp} to report UV luminosity limits which are one or two orders of magnitude lower than ours for targets at similar $z$, causing the \galex sample to perform worse than the \hst sample for $S\lesssim10$.

The choice of spectral model has a large effect on the results for the \galex sample but not for \hst. As Figure \ref{fig:Chev94FilterLum} shows, once the Lyman-$\alpha$ emission line enters the FUV band at $z\approx 0.1$, FUV luminosity dominates at higher redshifts compared to the other bands. The variation of this and several other emission lines introduces an additional redshift dependence in the line-emission model which is absent in the flat-spectrum model. As a result, the number of \sneia excluded by the line-emission model greatly increases above $S\approx10$ (i.e., PTF11kx-like events or stronger), to a maximum of 302 \sneia at $70\lesssim S\le 100$ in the top-left plot of Figure \ref{fig:ExcludedSNeIa}.

\subsection{Observational Constraints}
\label{ssec:ConstrainingCSM}

These results allow constraints to be placed on the occurrence rate of \sneia interacting with nearby CSM, \fcsm, at multiple epochs. Using a non-informative Jeffreys prior \citep{Jeffreys1946-BinomialPrior}, we estimate the 90\% binomial proportion confidence interval \citep[C.I.; see][]{Brown2001-BinomialStatistics} for \fcsm. Within a given range of $S$ and $t_\text{start}$ values, the number of excluded \sneia is the ``trials'' and the number of UV detections is the ``successes''.

Figure \ref{fig:CSMRateHist} presents the resulting upper bound of the 90\% C.I. for \fcsm for the \galex and \hst samples. The CSM model parameters are binned to the same intervals as in Figure \ref{fig:ExcludedSNeIa}, and the color scale is inverted to emphasize the inverse correspondence between the number of excluded \sneia and the upper limit on \fcsm. For the majority of our epochs we have no detections of CSM interaction, resulting in the lower bound of the 90\% C.I. being essentially zero for most parameter bins.

\begin{figure*}
    \centering
    \includegraphics[width=\textwidth]{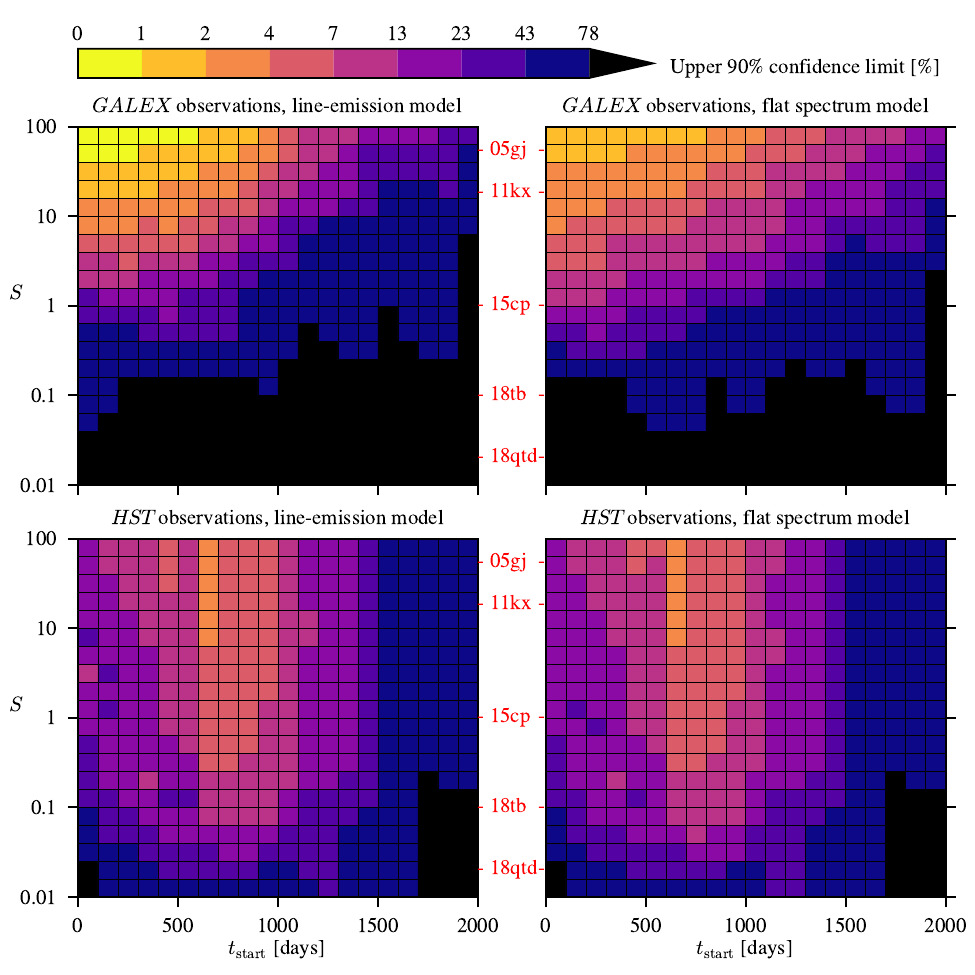}
    \caption{Upper limit of the 90\% confidence intervals on \fcsm, the occurrence rate of \sneia interacting with nearby CSM. Parameter bins are the same in Figure \ref{fig:ExcludedSNeIa}. \textit{Top:} upper limits from \galex non-detections, with zero detections in each bin. \textit{Bottom:} upper limits from \hst observations. Plots on the left use a line-emission model for SNe II \citep{Chevalier1994-TypeIICSM}, while plots on the right use a flat-spectrum model (see Section \ref{ssec:EmissionModel}). The relative scale factor is normalized so that unity corresponds to the UV luminosity of CSM interaction in SN 2015cp \citep{Graham2019-SN2015cp}. \galex data for the 50 days after discovery are omitted to avoid potential contamination from near-peak SN Ia flux, but model light curves are still injected down to $t_\text{start}=0$ days post-discovery.}
    \label{fig:CSMRateHist}
\end{figure*}

The \hst survey places tighter constraints on \fcsm below $S\approx3$, though this varies with $t_\text{start}$. Few \sneia were observed by \citet{Graham2019-SN2015cp} after $t_\text{disc}+1500$ days, limiting the effectiveness in that regime. By construction, the number of \sneia excluded from exhibiting CSM interaction is very similar for the line-emission and flat-spectrum models because the F275W luminosity is defined to be the same across both models for $S=1$ at the redshift of SN 2015cp \citep[$z=0.0413$,][]{Graham2019-SN2015cp}. Furthermore, because the \sneia observed in the \hst sample are all nearby ($\bar z\approx0.034$), emission lines do not move in and out of the filter. As the \hst limiting magnitudes were also similar for most of its targets, there is little dependence on $S$ where $S\ge1$. Positive detections increase both the lower and upper bounds of the 90\% C.I., so the \hst detections of ASASSN-15og at $+477$ days and SN 2015cp at $+664$ days result in an abrupt decrease in the upper 90\% C.I. at $t_\text{start}=600$ days as well as the isolated bins with tighter constraints between $0\le t_\text{start}\le 400$ days.

The constraints on \fcsm from the \galex sample depend mostly on $S$ and less on $t_\text{start}$. The large number of \sneia observed by \galex at $z\gtrsim0.1$ allows us to tightly constrain \fcsm for events one to two orders of magnitude more luminous than SN 2015cp. However, \fcsm is not well constrained for $S\lesssim3$, particularly at late times. There is also a stark difference between the line-emission and flat-spectrum models. As discussed in Section \ref{ssec:EmissionModel}, the effectiveness of the FUV band in the line-emission model increases dramatically for \sneia at $z\gtrsim0.1$, leading to very tight constraints for high $S$. By contrast, the upper 90\% C.I. for the flat-spectrum model decreases more smoothly with larger $S$ as it is driven mostly by sample size.

More stringent constraints on \fcsm can be obtained by analyzing the two surveys collectively. There is no overlap between targets in both samples, so we simply combine the number of \sneia excluded by each study to serve as the total number of binomial trials. As no \snecsm were detected by \galex, the number of successes is equal to the number of \hst detections.

Figure \ref{fig:CSMRateBounds} presents 90\% confidence intervals on the rate of CSM interaction among \sneia in the \galex and \hst samples alongside the combined sample (``All UV''). As before, confidence intervals are binned at 100-day increments of $t_\text{start}$. Each panel is a horizontal slice of Figure \ref{fig:CSMRateHist} and presents statistics for a single range of scale factors. We provide the results for scale factors of $S\approx 1$, $S\approx10$, and $S\approx100$, as these values roughly correspond to the strength of CSM interaction observed in SN 2015cp, PTF11kx, and SN 2005gj, respectively.

\begin{figure*}
    \centering
    \includegraphics[width=\textwidth]{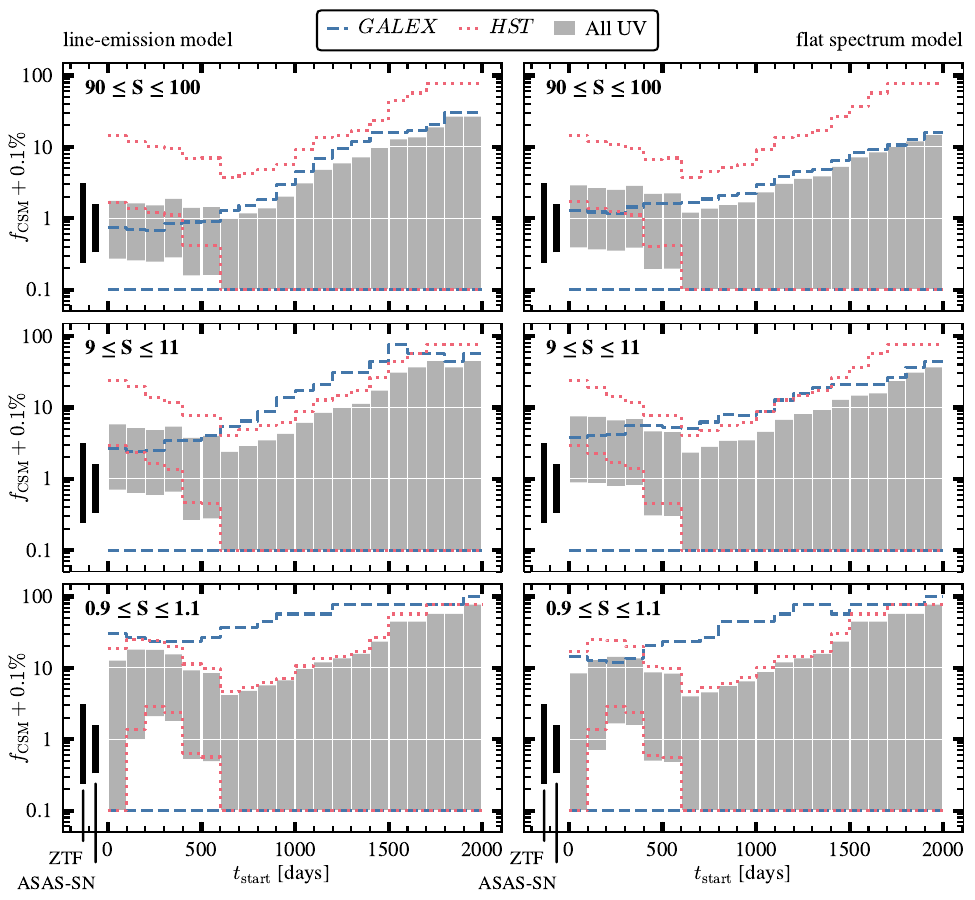}
    \caption{Constraints on \fcsm, the rate of CSM interaction among \sneia, at multiple epochs. Outlined and shaded regions represent 90\% binomial confidence intervals calculated from the number of excluded \sneia within each range of CSM interaction start times based on (red dotted lines) CSM detections and non-detections by \citet{Graham2019-SN2015cp}, (blue dashed lines) \galex non-detections, and (gray shaded region) both UV surveys combined. \textit{Top:} $90\le S\le100$, which constrains interactions on the order of SN 2005gj. \textit{Middle}: $9\le S\le11$, similar to PTF11kx. \textit{Bottom}: $0.9\le S\le1.1$, similar to SN 2015cp. The left panels use the \citet{Chevalier1994-TypeIICSM} line-emission model, while the right panels use a flat-spectrum model (see Section \ref{ssec:EmissionModel}). Confidence intervals based on early-time classifications in the ZTF \citep{Yao2019-ZTF2018} and ASAS-SN \citep{Holoien2017-ASASSN-DR1,Holoien2017-ASASSN-DR2,Holoien2017-ASASSN-DR3,Holoien2019-ASASSN-DR4} surveys are shown in black to the left of each plot at arbitrary $t_\text{start}$ values.}
    \label{fig:CSMRateBounds}
\end{figure*}

Included alongside our constraints on \fcsm are external constraints we derive from the All-Sky Automated Survey for SuperNovae \citep[ASAS-SN;][]{Shappee2014-ASASSN,Kochanek2017-ASASSN} and Zwicky Transient Facility \citep[ZTF;][]{Bellm2019-ZTF,Graham2019-ZTF} surveys. Over the first four years of operations, ASAS-SN observed three \snecsm out of 464 total \sneia \citep{Holoien2017-ASASSN-DR1,Holoien2017-ASASSN-DR2,Holoien2017-ASASSN-DR3,Holoien2019-ASASSN-DR4}, leading to a $90\%$ C.I. on the occurrence rate of $0.23\%\le f_\text{CSM}\le 1.51\%$. The ZTF 2018 sample \citep{Yao2019-ZTF2018} had only one SN Ia-CSM out of 127 \sneia, leading to $0.14\%\le f_\text{CSM}\le 3.04\%$. These crude estimates do not account for completeness in each survey, but considering \snecsm are typically overluminous compared to their normal counterparts, they are likely to be overrepresented in these catalogs.

Finally, we constrain \fcsm for broader ranges of $t_\text{start}$. Table \ref{tab:CSMRateBounds} presents upper 90\% confidence limits on the rate of late-onset CSM for several ranges of $t_\text{start}$ and $S$. We use only the line-emission spectral model because we consider it to be more representative of the true SED. For each given $t_\text{start}$ range we report the corresponding radius of the innermost shell of CSM $r_\text{CSM}$, assuming an ejecta velocity $v_\text{ej}\approx\SI{20000}{\kilo\meter\per\second}$ \citep{Garavini2005-SN1999ac}. We also report the eruption time $t_\text{erupt}$, in years before the SN Ia explosion, for material at $r_\text{CSM}$ assuming a shell expansion velocity $v_\text{exp}\approx\SI{100}{\kilo\meter\per\second}$, similar to PTF11kx \citep{Dilday2012-PTF11kx}.

\input{constraints.tex}

Strong interactions similar to that observed in SN 2005gj are rare: we constrain $f_\text{CSM}\lesssim1.6\%$ for $S\approx 100$ between $0-500$ days after discovery and $f_\text{CSM}\lesssim1\%$ between $500-1000$ days. We can also place tight constraints on the occurrence rate of PTF11kx-like events ($S\approx 10$), for which we find $f_\text{CSM}\lesssim 5.1\%$ between $0\le t_\text{start}\le 500$ days and $f_\text{CSM}\lesssim 2.7\%$ between $500\le t_\text{start}\le 1000$ days. \galex is much less effective at constraining \fcsm for SN 2015cp-like events ($S\approx 1$), but still manages a small improvement over the statistics from \hst alone with $f_\text{CSM}\lesssim 14\%$ between $0\le t_\text{start}\le 500$ days and $f_\text{CSM}\lesssim 4.8\%$ between $500\le t_\text{start}\le 1000$ days. This is consistent with the previous estimate of $f_\text{CSM}\lesssim 6\%$ for CSM within $r_\text{CSM}\approx\SI{3e17}{\centi\meter}$ reached by \citet{Graham2019-SN2015cp}. Finally, for events on the scale of ASASSN-18tb ($S\approx 0.1$), we constrain $f_\text{CSM}\lesssim 16\%$ between $0\le t_\text{start}\le 500$ days and $f_\text{CSM}\lesssim8.6\%$ between $500\le t_\text{start}\le1000$ days almost exclusively from the \hst data. These results represent the most thorough attempt thus far to constrain \fcsm for an unbiased sample of \sneia. 

%****************************************************************************

\section{Conclusions}
\label{sec:Conclusions}

We present results from our search for late-onset CSM interaction among \sneia. \galex serendipitously observed \samplesize \sneia at $z<0.5$ both before and after discovery. Four \sneia (SNe 2007on, 2008hv, 2009gf, and 2010ai) are detected near-peak in the \galex NUV filter but no evidence of \sneia interacting with a nearby CSM was found. 

With the UV non-detections of \injrecsize \sneia, we implement an injection-recovery procedure to estimate the intrinsic fraction of \sneia interacting with CSM, \fcsm. Due to the lack of models in the literature addressing the UV emission and light curve evolution of \snecsm, we make several simple assumptions about the underlying SED and its temporal evolution. Combining our \galex observations with the \hst survey performed by \citet{Graham2019-SN2015cp}, we can constrain \fcsm for a broad range of scale factors $S$ relative to SN 2015cp (a proxy for the CSM luminosity) and times when the SN ejecta first encounters the CSM $t_\text{start}$.

We strongly constrain the most luminous events, such as those similar to SN~2005gj ($S\approx100$, or $L_{\text{H}\alpha}\sim\SI{e41}{\erg\per\second}$), at high confidence with $f_\text{CSM}\lesssim1.6\%$ between $0-500$ days after discovery and $f_\text{CSM}\lesssim1\%$ between $500-1000$ days. Moderate-luminosity CSM interactions similar to that seen in PTF11kx ($S\approx10$, or $L_{\text{H}\alpha}\sim\SI{e40}{\erg\per\second}$) are still rare, with $f_\text{CSM}\lesssim5.1\%$ and $f_\text{CSM}\lesssim2.7\%$, respectively, for the same time scales. SN~2015cp-like events ($S\approx1$, or $L_{\text{H}\alpha}\sim\SI{e39}{\erg\per\second}$) are constrained to $f_\text{CSM}\lesssim4.8\%$ between $500-1000$ days with weaker constraints at other time scales. For the weakest CSM interactions (e.g., ASASSN-18tb), our observations do not place meaningful constraints, highlighting the need for further monitoring of \sneia out to late epochs, especially in the UV where CSM is easy to distinguish from the underlying ejecta emission.

Finally, this study reinforces the need for consistent monitoring of \sneia at late times. Since observations of most \sneia last for just a few months after the explosion, any instance of late-onset CSM interaction is likely to be systematically missed. In addition, the ability to constrain \fcsm is limited by the lack of models for \snecsm in the UV. As \snecsm potentially originate through the SD progenitor channel, our constraints may inform future studies on the nature of SN Ia progenitors.

%****************************************************************************

\acknowledgements

\software{\gphoton \citep{Million2016-gPhoton}, AstroPy \citep{Astropy2013,Astropy2018}, astroquery \citep{Ginsburg2019-astroquery}, statsmodels \citep{seabold2010statsmodels}, SciPy \citep{2020SciPy-NMeth}, NumPy \citep{2020NumPy-Array}, Matplotlib \citep{Hunter2007-matplotlib}, and pandas \citep{mckinney-proc-scipy-2010,reback2020pandas}.}

We thank Christopher Kochanek and Connor Auge for providing useful comments on the manuscript. We thank Chase Million and Scott Fleming for useful discussions about \gphoton. We also thank Greg Aldering, Melissa Graham, and David Sand for their help searching for archival classification spectra.

LOD acknowledges support from Research Experience for Undergraduates program at the Institute for Astronomy, University of Hawaii-Manoa funded through NSF grant 6104374. LOD would like to thank the Institute for Astronomy for their kind hospitality during the course of this project. MAT acknowledges support from the DOE CSGF through grant DE-SC0019323. BJS is supported by NSF grants AST-1907570, AST-1920392, and AST-1911074.

This research has made use of the SVO Filter Profile Service\footnote{\url{http://svo2.cab.inta-csic.es/theory/fps3/}} supported from the Spanish MINECO through grant AYA2017-84089. This research also makes use of the NASA/IPAC Extragalactic Database (NED), which is funded by the National Aeronautics and Space Administration and operated by the California Institute of Technology.

%****************************************************************************

\bibliography{references}
\bibliographystyle{aasjournal}

%****************************************************************************

\appendix

\section{New Redshift Measurements}\label{app:manualz}

\input{newz.tex}

For \sneia with redshifts given to $< 2$ decimal places, we attempt to manually improve the redshift determination based on available data. The new redshifts are provided in Table \ref{tab:manualz} including the catalog redshift, the redshift derived in this work, the spectrum data source, and the method used to measure the redshift and associated uncertainty. SNe 2009cp and 2009cu have publicly-available reduced spectra in the Weizmann Interactive Supernova Data Repository (WISeREP, \citealp{yaron2012}) whereas SNe 2009kt, 2009kv, and 2009kx have Gemini Multi-Object Spectrograph \citep[GMOS; ][]{hook2004} data available through the Gemini Observatory Archive. The raw GMOS spectra were reduced using the GMOS Data Reduction Cookbook\footnote{\url{http://ast.noao.edu/sites/default/files/GMOS_Cookbook/}} with calibration frames taken near the time of observation. 

The spectrum of SN~2009kt shows host-galaxy emission lines evident in the extracted spectrum, and the host-galaxy origin is confirmed via extended emission in the 2D spectra. Emission lines from \Halpha, H$\beta$, the [OIII] 4959/5007\AA{} doublet, and the [SII] 6716/6731\AA{} doublet are observed, and we fit these lines simultaneously to derive the host-galaxy redshift. 

The remaining SNe Ia do not have host-galaxy lines in their spectra so we estimate a redshift using the SuperNova IDentification code \citep[SNID; ][]{Blondin2007-SNID}. After confirming the best matches to the observed spectrum are SNe Ia, we restrict the correlation templates to only SNe Ia and use the 10 best matches to estimate the redshift and its uncertainty. 

\section{Photometric Precision and Stability}
\label{app:Photometry}

\galex photometry has been used previously in co-added images \citep[e.g.,][]{leroy2019, bracco2020} and to study short-term intra-visit stellar variability \citep[e.g.,][]{boudreaux2017, tucker2018, rowan2019} but never to our knowledge used for long-term monitoring as we do here. Thus, we want to include as much photometry as possible without sacrificing photometric quality and stability. To these ends, we run several photometric tests to validate our assumptions on the \galex photometry.

The \textsc{detrad} column output by \gphoton denotes the average photon event distance from detector center. The ``detector edge'' flag is generated when any photon event occurs at $> 0.5~\rm{deg}$ from detector center, yet the \galex detectors have radii of $\approx 0.62~\rm{deg}$. Photometry near the edge of the detectors is untrustworthy as the detector response at the edge is poorly characterized \citep{Morrissey2007-GALEXcalib} and has reduced count rates due to the \galex dithering process. However, the $0.5~\rm{deg}$ cut is likely too conservative for our purposes so we explore using a larger maximum radius to improve our photometric completeness. 

Figure \ref{fig:detrad} provides the difference between the MCAT reference magnitudes and the \textsc{gAperture}-derived magnitudes for an aperture radius of 6$^{\prime\prime}$ (the \galex MCAT \textsc{aper4} radius). We see there is little difference between MCAT and \textsc{gAperture} until $\sim 0.6~\rm{deg}$ from detector center. To ensure photometric quality we use an updated \textsc{detrad} cut at $0.6~\rm{deg}$, increasing our effective \galex detector size by $\sim 45\%$.

\begin{figure}[bth]
    \centering
    \plotone{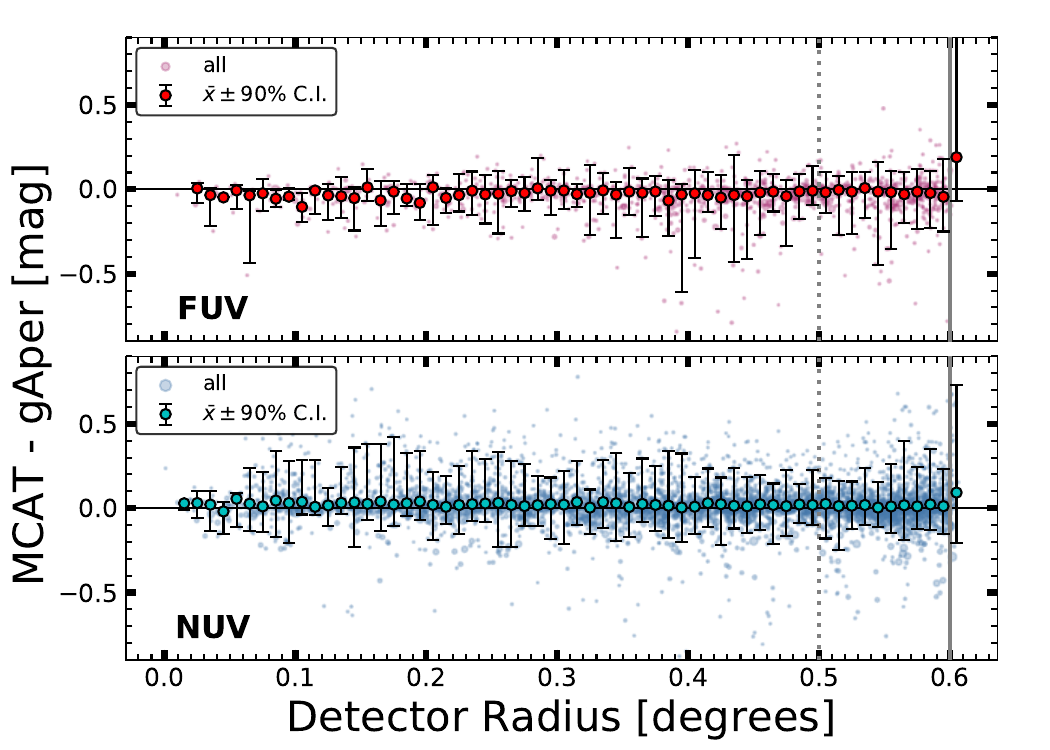}
    \caption{Difference between MCAT magnitudes and \textsc{gAperture} magnitudes as a function of average detector radius. Individual point sizes are inversely proportional to their respective uncertainties. Bold points mark the median and 90\% confidence interval for equally-spaced 0.02~deg-wide bins. The dotted gray line signifies the nominal \gphoton detector edge flag set at $> 0.5~\rm{deg}$ from center and the solid gray line signifies our updated cut at $> 0.6~\rm{deg}$.}
    \label{fig:detrad}
\end{figure}

\section{Host-Galaxy Systematic Uncertainty}
\label{app:host-systematics}

A key component of our analysis is requiring \galex observations both before and after the SN Ia discovery date so the host-galaxy flux can be effectively removed. \sneia with $\geq 5$ host-galaxy observations have sufficient data to both estimate the mean host-galaxy flux and the ensuing uncertainty (see Section \ref{subsec:hostgal-subtraction}). However, for \sneia host galaxies with $< 5$ observations, we risk under-estimating the true uncertainty of the host-galaxy flux and thus under-estimating the ensuing host-subtracted SN fluxes. Therefore, we calculate a systematic uncertainty for host galaxies with few observations to ensure our results are statistically robust.

Figure \ref{fig:hostsys-magdep} compares the reference MCAT magnitude to the single-epoch \gphoton aperture photometry magnitude for a given MCAT source. This provides a rough estimate of the systematic uncertainty as a function of UV brightness. We note that the NUV suffers from higher scatter due to increased source crowding even though the FUV typically has fewer detected photons \citep{Million2016-gPhoton}. We implement adaptive binning when computing the bin size, requiring 100 stars per bin. This approach prevents brighter bins from having very few objects per bin and retains roughly equal bin sizes for higher magnitudes. For each bin, we apply iterative sigma clipping then compute the weighted mean and standard deviation shown as the solid colored lines in Figure \ref{fig:hostsys-magdep}. We include an anchor at the bright end of the distribution from \citet{Morrissey2007-GALEXcalib} of $\Delta m = \pm 0.03~(0.05)$ mag for the NUV (FUV) data, respectively. We approximate the systematic uncertainty with a power-law function, $\Delta m_{\rm{sys}} = A(\frac{m}{1~\rm{mag}}-14)^B$, where $A$ and $B$ are fitted coefficients and $m$ is the filter-specific \galex magnitude. To reduce the covariance in the fitting process, $m$ is offset by 14~mag. We find $(A_{\rm{NUV}} = (4.94\pm0.83)\times10^{-7}~ \textrm{mag},~ B_{\rm{NUV}} = 6.17\pm0.08)$ and $(A_{\rm{FUV}} = (4.78\pm1.61)\times10^{-4}~ \textrm{mag},~ B_{\rm{FUV}} = 2.60\pm0.16)$. These are rough approximations but should suffice for our purposes of preventing an underestimate of the host-galaxy flux.

\begin{figure}[tb]
    \centering
    \plotone{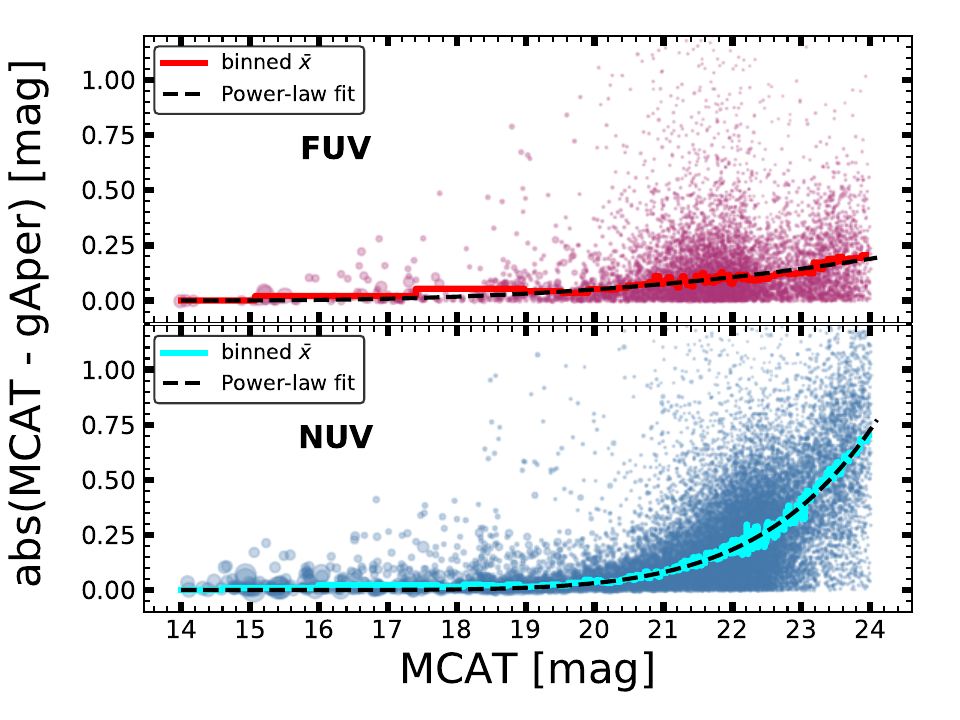}
    \caption{Difference between MCAT and \textsc{gAperture} magnitudes as a function of MCAT magnitude. Colored points are individual sources with the point size proportional to $1/\sigma$ (i.e., smaller point = larger uncertainty). Solid colored lines are the binned weighted mean for each filter using adaptive bin sizes of 100 sources (see text). Dashed lines represent simple power-law fits to the binned data.}
    \label{fig:hostsys-magdep}
\end{figure}

\section{False Positives}
\label{app:false-detect}

ESSENCEn263 has significant NUV detections between 1421 and 1820 days after discovery, with a maximum $11.9\sigma$ detection relative to the host flux as shown in Figure \ref{fig:ESSENCEn263-lc}. However, the host-galaxy is a known broad-line AGN \citep[SDSS ObjID = 1237679253596340445;][]{2016SDSSD.C...0000:} and offset from the position of ESSENCEn263 by $\sim 2.6^{\prime\prime}$ but within our $6^{\prime\prime}$-radius aperture. Figure \ref{fig:ESSENCEn263-images} presents \textsc{gMap} images for ESSENCEn263, which confirm the NUV excess is centered on the host-galaxy and not the SN Ia location.

\begin{figure}[tb]
    \centering
    \plotone{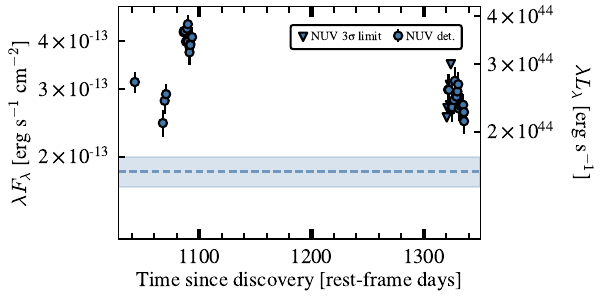}
    \caption{\galex NUV light curve of ESSENCEn263, a likely AGN. The blue points represent $\ge 3\sigma$ detections and the inverted triangles signify $3\sigma$ non-detection limits. The dashed line and shaded region represent the host-galaxy flux and associated $1\sigma$ uncertainty, respectively. The vertical axis on the right-hand side converts observed flux $\lambda F_\lambda$ to luminosity $\lambda L_\text{UV}$, corrected for Milky Way extinction.}
    \label{fig:ESSENCEn263-lc}
\end{figure}

\begin{figure}[tb]
    \centering
    \plotone{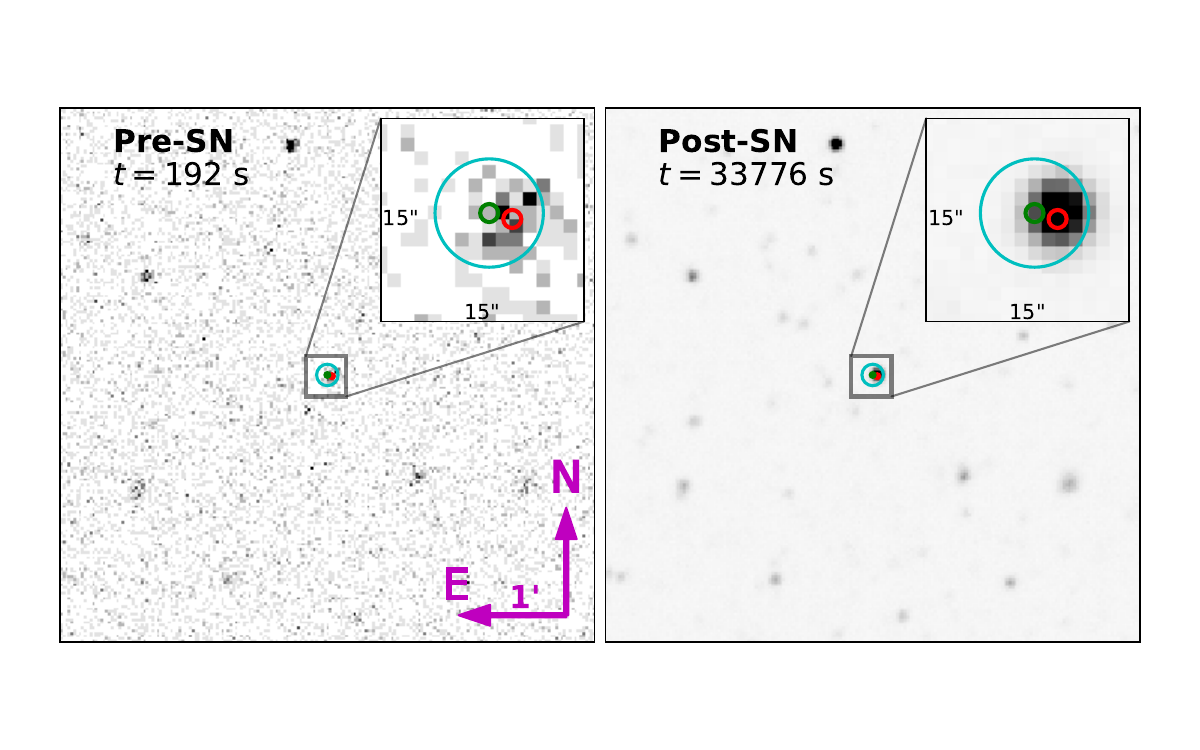}
    \caption{Pre- (left) and post-SN (right) \galex NUV images of ESSENCEn263. The red and green circles have radii of $1^{\prime\prime}$ each and are centered on the host-galaxy \citep[from SDSS;][]{2016SDSSD.C...0000:} and the SN Ia \citep[from ESSENCE;][]{Miknaitis2007-ESSENCE}, respectively. The cyan circle represents the $6^{\prime\prime}$-radius photometric aperture. The post-discovery detections are consistent with the location of the host-galaxy, a known AGN, instead of the SN Ia. All images and insets have been corrected for exposure time and are on the same scale.}
    \label{fig:ESSENCEn263-images}
\end{figure}

\end{document}

%% file: short_sample.tex
\movetabledown=2in
\begin{rotatetable*}
\newcolumntype{z}{>{\raggedright\arraybackslash}p{1.75in}}
\begin{deluxetable*}{lcllrRrrDr@{$\pm$}llz}
\tablecaption{\normalsize Basic information for our sample of \sneia.
\label{tab:Targets}}
\tablewidth{0pt}
\tabletypesize{\footnotesize}
\tablehead{
\colhead{Target Name} & \colhead{Disc. Date} & \colhead{R.A.} & \colhead{Dec.} &  
\colhead{Obs.\tablenotemark{1}} & \colhead{$t_\text{first}$\tablenotemark{2}} & 
\colhead{$t_\text{last}$\tablenotemark{3}} & \colhead{$t_\text{next}$\tablenotemark{4}} & 
\multicolumn2c{Redshift} & \multicolumn2c{Distance} & \colhead{$A_V$\tablenotemark{5}} & \colhead{Reference(s)}
\vspace{-6pt} \\
\colhead{} & \colhead{} & \colhead{[h:m:s]} & \colhead{[d:m:s]} &
\colhead{} & \colhead{[days]} & 
\colhead{[days]} & \colhead{[days]} & 
\multicolumn2c{} & \multicolumn2c{[Mpc]} & \colhead{[mag]} & \colhead{}
}
\decimals
\startdata
ESSENCEg097 &  2004-10-05 &   $23:27:37.16$ &       $-09:35:21$ &            23 &           -407 &          2181 &           357 &    0.343 &       1464 &            103 &  0.086 &                                          \citet{Miknaitis2007-ESSENCE} \\
ESSENCEg142 &  2004-10-09 &    $23:28:37.7$ &       $-08:45:04$ &            16 &            -35 &          2177 &           353 &    0.404 &       1725 &            121 &  0.078 &                                          \citet{Miknaitis2007-ESSENCE} \\
ESSENCEg230 &  2004-10-17 &   $01:11:56.31$ &     $+00:07:27.7$ &            19 &           -376 &          2532 &          1081 &    0.392 &       1674 &            117 &  0.084 &                                          \citet{Miknaitis2007-ESSENCE} \\
ESSENCEm027 &  2005-09-26 &   $01:09:15.01$ &     $+00:08:14.8$ &            14 &           -720 &          1870 &           742 &    0.289 &       1233 &             86 &  0.078 &                                          \citet{Miknaitis2007-ESSENCE} \\
ESSENCEm043 &  2005-09-26 &   $23:29:51.73$ &     $-08:56:46.1$ &            16 &           -387 &          1825 &             1 &    0.266 &       1134 &             80 &  0.080 &                                          \citet{Miknaitis2007-ESSENCE} \\
ESSENCEm062 &  2005-09-25 &  $01:09:52.902$ &    $+00:36:19.03$ &            14 &           -719 &          1829 &           729 &    0.314 &       1340 &             94 &  0.067 &                                          \citet{Miknaitis2007-ESSENCE} \\
ESSENCEm075 &  2005-09-26 &   $23:24:42.29$ &     $-08:29:08.7$ &            16 &           -708 &          2209 &           360 &    0.102 &        423 &             30 &  0.099 &                                          \citet{Miknaitis2007-ESSENCE} \\
ESSENCEn263 &  2005-11-22 &   $02:05:14.95$ &     $-04:56:39.1$ &            40 &           -410 &          1820 &          1421 &  0.36264 &       1549 &            109 &  0.070 &                                          \citet{2016SDSSD.C...0000:} \\
ESSENCEn278 &  2005-11-24 &   $23:28:17.55$ &     $-09:23:12.4$ &            22 &           -446 &          1766 &           296 &    0.304 &       1297 &             91 &  0.109 &                                          \citet{Miknaitis2007-ESSENCE} \\
ESSENCEn326 &  2005-11-24 &   $23:29:58.59$ &     $-08:53:12.5$ &            16 &           -446 &          1766 &           291 &  0.26316 &       1128 &             79 &  0.082 &  \citet{2007SDSS6.C...0000:,2004SDSS3.C...0000:,2007ApJ...660..239K} \\
ESSENCEn400 &  2005-11-26 &   $01:13:13.26$ &     $-00:23:25.9$ &            16 &           -781 &          1767 &           676 &    0.424 &       1811 &            127 &  0.085 &                                          \citet{Miknaitis2007-ESSENCE} \\
ESSENCEp425 &  2005-11-24 &   $23:29:56.19$ &     $-08:34:24.3$ &            20 &           -767 &          1766 &           291 &    0.458 &       1956 &            137 &  0.085 &                                          \citet{Miknaitis2007-ESSENCE} \\
ESSENCEp434 &  2005-11-24 &   $01:12:40.25$ &     $+00:14:56.6$ &            15 &           -779 &          2129 &           683 &    0.339 &       1447 &            101 &  0.094 &                                          \citet{Miknaitis2007-ESSENCE} \\
ESSENCEq002 &  2006-09-16 &  $02:05:12.945$ &   $-03:39:00.723$ &            15 &          -1029 &          1546 &            64 &   0.3469 &       1482 &            104 &  0.064 &                                          \citet{narayan2016} \\
ESSENCEq022 &  2006-09-11 &  $01:12:03.864$ &  $-00:01:28.9452$ &            19 &          -1070 &          1838 &           387 &  0.22637 &        965 &             68 &  0.083 &                      \citet{2007SDSS6.C...0000:,2016SDSSD.C...0000:} \\
ESSENCEr185 &  2006-10-31 &  $01:11:48.245$ &    $-00:29:49.46$ &            12 &          -1120 &          1428 &           337 &  0.18011 &        767 &             54 &  0.066 &                      \citet{2007SDSS6.C...0000:,2016SDSSD.C...0000:} \\
ESSENCEr317 &  2006-11-14 &  $01:13:24.658$ &   $+00:51:27.757$ &            28 &           -778 &          1774 &           312 &   0.3361 &       1435 &            101 &  0.070 &                                          \citet{narayan2016} \\
Hawk        &  2004-11-05 &   $12:35:41.16$ &    $+62:11:37.19$ &           211 &           -281 &          2315 &           163 &  0.49673 &       2129 &            149 &  0.031 &                    \citet{2011AandA...528A..35M,2004AJ....127.3121W} \\
HST04Sas    &  2004-05-23 &  $12:36:54.125$ &    $+62:08:22.21$ &           213 &           -118 &          2481 &           329 &  0.44643 &       1914 &            134 &  0.031 &                      \citet{2004ApJ...617..240K,2004AJ....127.3121W} \\
\enddata
\tablenotetext{1}{Number of \galex observations in both bands}
\tablenotetext{2}{Number of days between the first \galex observation in either band and the discovery date}
\tablenotetext{3}{Number of days between the discovery date and the last \galex observation in either band}
\tablenotetext{4}{Number of days between the discovery date and the next \galex observation in either band}
\tablenotetext{5}{$V$-band galactic extinction}
% \tablenotetext{a}{Offset between NED entry and \galex sky coordinates is larger than 30 kpc}
\tablecomments{Table~\ref{tab:nearby-historical} is published in its entirety in the electronic 
edition of the {\it Astrophysical Journal}.  A portion is shown here 
for guidance regarding its form and content.}
\end{deluxetable*}
\end{rotatetable*}

%% file: nearby_historical.tex
\begin{deluxetable*}{llllll}
\tablecaption{Supernovae observed by GALEX only after discovery\label{tab:nearby-historical}}
\tablehead{
\colhead{Name} & \colhead{Date} & \colhead{NUV} &
\colhead{FUV} & \colhead{$t_{\rm first}$} & \colhead{$t_{\rm last}$} \\
\colhead{} & \colhead{(YYYY-MM-DD)} & \colhead{[\#]} & \colhead{[\#]} &
\colhead{[days]} & \colhead{[days]}
}
\startdata
SN1937D& 1937-09-09& 0&  1&25295&25295 \\
SN1954B& 1954-04-27& 3&  3&17952&19342 \\
SN1957A& 1957-02-26&14& 17&17128&19016 \\
SN1960F& 1960-04-17& 0&  4&17167&18278 \\
SN1960H& 1960-06-18& 2&  2&15984&16690 \\
\enddata
\tablecomments{Table~\ref{tab:nearby-historical} is published in its entirety in the electronic 
edition of the {\it Astrophysical Journal}.  A portion is shown here 
for guidance regarding its form and content.}
\end{deluxetable*}

%% file: scale_factors.tex
\begin{deluxetable*}{lDDDl}
\tablecaption{\normalsize Scale factors for reference \snecsm from \Halpha line luminosity ratios.\label{tab:ReferenceScaleFactors}}
\tabletypesize{\small}
\tablewidth{0pt}
\tablehead{
    \colhead{SN} & \multicolumn2c{$L_{\text{H}\alpha}$} & \multicolumn2c{$S$} & \multicolumn2c{Epoch} & \colhead{Source}
    \vspace{-2pt} \\
    \colhead{} & \multicolumn2c{[$10^{39}$ erg/s]} & \multicolumn2c{} & \multicolumn2c{[days]} & \colhead{}
}
\decimals
\startdata
SN 2005gj   & 118.  & 54.   & +111 & \citet{Prieto2007-SN2005gj}     \\
PTF11kx     & 40.6  & 19.   & +371  & \citet{Silverman2013-PTF11kx}   \\
SN 2015cp   & 2.2   & 1     & +694  & \citet{Graham2019-SN2015cp}     \\
ASASSN-18tb/SN 2018fhw      & 0.22  & 0.10  & +139  & \citet{Kollmeier2019-SN2018fhw} \\
ATLAS18qtd/SN 2018cqj       & 0.038 & 0.02  & +193  & \citet{Prieto2020-SN2018cqj}    \\
\enddata
\end{deluxetable*}

%% file: constraints.tex
\begin{deluxetable*}{LLDDDLL}
\tablecaption{\normalsize Upper 90\% confidence limit on CSM interaction rate using the line-emission model.\label{tab:CSMRateBounds}}
\tablewidth{0pt}
\tabletypesize{\small}
\tablehead{
    \colhead{$S$} & \colhead{$t_\text{start}$ [days]} & \multicolumn2c{\galex [\%]} & \multicolumn2c{\hst [\%]} & \multicolumn2c{All UV [\%]} & \colhead{$r_\text{CSM}$\tablenotemark{a} [\SI{e16}{\centi\meter}]} & \colhead{$t_\text{erupt}$\tablenotemark{b} [yr]}
}
\decimals
\startdata
% scale factor      t_start     GALEX   HST     All     r_csm   t_erupt
90-100              & 0-500     & 0.65  & 11.   & 1.6   & 0-9   & 0-270     \\
(\sim\text{05gj})   & 500-1000  & 1.4   & 4.2   & 1.0   & 9-17  & 270-540   \\
                    & 1000-1500 & 7.6   & 13.   & 5.0   & 17-26 & 540-820   \\
                    & 1500-2000 & 21.   & 57.   & 17.   & 26-35 & 820-1100  \\
\hline
9-11                & 0-500     & 2.7   & 14.   & 5.1   & 0-9   & 0-270     \\
(\sim\text{11kx})   & 500-1000  & 6.2   & 4.7   & 2.7   & 9-17  & 270-540   \\
                    & 1000-1500 & 23.   & 13.   & 9.0   & 17-26 & 540-820   \\
                    & 1500-2000 & 57.   & 57.   & 36.   & 26-35 & 820-1100  \\
\hline
0.9-1.1             & 0-500     & 23.   & 18.   & 14.   & 0-9   & 0-270     \\
(\sim\text{15cp})   & 500-1000  & 31.   & 5.5   & 4.8   & 9-17  & 270-540   \\
                    & 1000-1500 & 77.   & 15.   & 13.   & 17-26 & 540-820   \\
                    & 1500-2000 & 77.   & 57.   & 44.   & 26-35 & 820-1100  \\
\hline
0.1-0.2             & 0-500     & 77.       & 17.   & 16.   & 0-9   & 0-270    \\
(\sim\text{18tb})   & 500-1000  & 77.       & 9.0   & 8.6   & 9-17  & 270-540   \\
                    & 1000-1500 & \nodata\tablenotemark{c}   & 21.   & 21.   & 17-26 & 540-820   \\
                    & 1500-2000 & \nodata   & 77.   & 77.   & 26-35 & 820-1100  \\
\enddata
\tablenotetext{a}{Corresponding radius of the innermost shell of CSM assuming an ejecta velocity $v_\text{ej}\approx\SI{20000}{\kilo\meter\per\second}$ \citep{Garavini2005-SN1999ac}.}
\tablenotetext{b}{Eruption time in years before the SN Ia explosion for material at $r_\text{CSM}$, assuming a shell expansion velocity $v_\text{exp}\approx\SI{100}{\kilo\meter\per\second}$ \citep{Dilday2012-PTF11kx}.}
\tablenotetext{c}{Indicates no \sneia were excluded by \galex observations in this range.}
\end{deluxetable*}

%% file: newz.tex
\begin{deluxetable}{ccccc}
\tablecaption{SNe Ia with updated redshifts derived in this work.
\label{tab:manualz}}
\tablehead{
    \colhead{Name} & \colhead{Catalog Redshift} & \colhead{New Redshift} & \colhead{Data Source} & \colhead{Measurement Type}
}
\startdata
SN~2009cp & 0.22 & $0.225\pm0.008$ & WISeREP & SNID \\
SN~2009cu & 0.10 & $0.099\pm0.003$ & WISeREP & SNID \\
SN~2009kt & 0.27 & $0.27592\pm0.00005$ & Gemini & host-galaxy lines \\
SN~2009kv & 0.32 & $0.315\pm0.008$ & Gemini & SNID \\
SN~2009kx & 0.23 & $0.217\pm0.003$ & Gemini & SNID \\
\enddata
\end{deluxetable}